# Giant enhancement of the NMR resolution and sensitivity of the critical solutions of hyperpolarized fluids within closed carbon nanotubes

M.G. Rudavets*

*Institute of Problems of Chemical Physics, RAS, Chernogolovka, Moscow region, 142432, Russia*

We predict the effect of the nuclear spin density fluctuations on the NMR spectra of the critical solutions of spin-carrying atoms in the closed carbon nanotubes (CNTs). The paper is mostly concerned with the critical binary isotopic solution $^{129}Xe - ^{132}Xe$ in the CNTs. The effective dipolar coupling $\bar{v}$ of the nuclear spin-carrying $^{129}Xe$ nanofluid is shown to separated into two terms, namely the term $\bar{v}_0$ of the non-correlated $^{129}Xe$ atoms [J.Baugh et al., Science **294**, 1505 (2001)] and the additional term $\bar{v}_\Delta$ depending on the correlation function of the density fluctuations of the $Xe$ atoms. Whereas the coupling $\bar{v}_0$ falls off to zero as $\Omega^{-1}$ with increasing the volume $\Omega$ of the CNTs, the correlation term $\bar{v}_\Delta$ remains finite for the long CNTs containing the critical solution of the $^{129}Xe$ atoms. This correlation term behaves as $\bar{v}_\Delta \sim \left( (A\xi_0)^2 \bar{n}_t \right)^{-1} \Phi(\tau)$ where $\xi_0$ denotes the correlation length of the solution far from the critical point (CP), the value $\bar{n}_t$ refers to the total density of the $Xe$ atoms at the CP of the nanofluid with the cross section area $A$, and the bell-shaped function $\Phi(\tau)$ of the reduced temperature $\tau = |T-T_c|/T_c$ is derived within the Landau-Ginzburg framework. When the temperature $T$ is approached to the critical temperature $T_c$, the residual dipolar coupling tends to the finite value $\bar{v}_\Delta \approx 10$ Hz for $^{129}Xe$ fluid in the closed long tubes. As $T \gg T_c$, the function $\Phi(\tau)$ and the coupling $\bar{v}_\Delta$ tend to zero giving the conventional result $\bar{v} = \bar{v}_0$ for the average dipolar coupling. In order to achieve the abnormally large effective coupling $\bar{v}_\Delta \gg \bar{v}_0$ the following three conditions should be met: (1) the large mobility of $^{129}Xe$ atoms, (2) the maximal isothermal compressibility of the $Xe$ nanofluid within the CNTs that have to be (3) long and closed. We discuss three applications of such a behavior of the effective dipolar coupling $\bar{v}$. First, when $|\tau| \to 0$, the Fourier transform of the free induction decay $\langle I_x \rangle(\omega)$, being the Gaussian envelope of the $N$ resonant peaks from the $N$ magnetically equivalent $^{129}Xe$ atoms, is broadened so wide that the function $\langle I_x \rangle(\omega)$ splits into the lattice of the $N$ equidistant resonances with the finite spacing $3\bar{v}_\Delta$. Second, the absorption line shape of the critical hyperpolarized $^{129}Xe$ in the spin state $I = N/2, m = -N/2$ has single peak $\chi''(\omega) \sim N\delta(\omega + \omega_*)$ at the frequency $\omega_* = \omega_0 - \frac{3}{2}\bar{v}_\Delta (N-1)$ with the large deviation from the Larmor frequency $|\omega_0| = |\gamma_{129_{Xe}}| B_0$ for $N \gg 1$. The dipolar field of the critical hyperpolarized nanofluid in the spin state $I = N/2$, $m = N/2$ inverts the total magnetic field if $N \geq 1 + \frac{2}{3} |\gamma_{129_{Xe}}| B_0 / \bar{v}_\Delta$. Third, we discuss the spontaneous superradiation of the nanofluid in the course of the depolarization $\langle I_x \rangle(t)$ in the low-field resonator specified by the Larmor frequency $|\omega_0| \ll \bar{v}\sqrt{N}$. At the CP of the nanofluid, the depolarization $\langle I_x \rangle(t)$ causes the bursts of the dissipated power $P \sim \bar{v}_\Delta^2 N^3$ in the pick-up coil. In the opposite limiting case $|\omega_0| \gg \bar{v}\sqrt{N}$, the depolarization $\langle I_x \rangle(t)$ has the Dicke's power $P_D \sim \omega_0^2 N^2$. Far from the CP of the nanofluid, the dissipated power $P$ scales linear with $N$ for the fixed density $\bar{n} = N/\Omega$.

# I. INTRODUCTION

Successful synthesis of carbon nanotubes[1,2] (CNTs) and the great promise of CNT technology[3] have created an explosion of interest[4,5] in the properties of the fluids within the CNTs with the goal of constructing the functionalized nanodevices. The nanofluids are becoming a useful probe in detecting the hollow space of the nanotubes and CNTs membranes,[6-8] synthetic pores,[9-11] biological channels[12,13] as well as critical for controlling on the molecular level the implementation of nanofluidic devices,[6,14-16] nanojets, nanopipets,[17,18] as a means for pure hydrogen storage for fuel cell,[19] potentially interesting for nano-medicine[20] as nano-arrays for sensing the chemical agents in nano-reactors, single cell analysis, biomolecular sorting,[21,22] drug delivery,[23,24] etc. The filling of the CNTs not only does create the new composite systems but, in turn, confinement impose the finite-size effects on the nanofluid modifying the transport, thermodynamic, spectroscopic properties of the nanofluid relative to those found in the bulk. Moreover the confinement allows for studying the fluctuation and coherence effects of the nanofluids aimed at constructing nanodevices with maximal efficiency.

We want to show that to achive an enhancement of the intensities and the resonance frequency shifts of the NMR signals the nanofluid should possess the coherence both over the spatial degrees of freedom at the thermodynamic critical point (CP) and over the nuclear spin states of the hyperpolarized nanofluid. Recently the NMR spectroscopy has been used for exploring the inner space of the pores due to the relationship between the inhomogeneous NMR line of the nanofluids and the shape of the pores.[25] Consistent description of how the confinement affect the NMR line-shape was found relying on the picture of non-interacting spin carrying atoms of the nanofluid. Although experimental evidence indicates that this approximation is appropriate inasmuch as the NMR line shape is concerned,[25,26] a more consistent treatment requires including the correlations between the atoms of the nanofluid. In close parallel with Ref. 25, a general microscopic consideration based on the adiabatic theory shows (Section III.A) that for the nuclear spin nanofluid with the dipolar interaction $v(\vec{r}_1, \vec{r}_2)$ between magnetically equivalent nuclear spins, the average interatomic dipole-dipole interaction is specified by the total dipolar energy per pair of the nuclear spins,

$$\bar{v} = \frac{1}{N^2} \iint_\Omega d^3\vec{r}_1 d^3\vec{r}_2 \ v(\vec{r}_1, \vec{r}_2) \langle n(\vec{r}_1) n(\vec{r}_2) \rangle = \bar{v}_0 + \bar{v}_\Delta \qquad (1.1)$$

In Eq. (1.1), the average dipolar interaction $\bar{v}$ is discriminated into the two terms stemming from the dipolar interactions of the independent atoms and the dipolar interactions of the correlated atoms, respectively,

$$\bar{v}_0 = \frac{1}{\Omega^2} \iint_\Omega d^3\vec{r}_1 d^3\vec{r} \ v(\vec{r}_1, \vec{r}_2) \qquad (1.2)$$

$$\bar{v}_\Delta = \frac{1}{N^2} \iint_\Omega d^3\vec{r}_1 d^3\vec{r}_2 \ v(\vec{r}_1, \vec{r}_2) \langle \delta n(\vec{r}_1) \delta n(\vec{r}_2) \rangle \qquad (1.3)$$

The coupling $\bar{v}_0$ of Eq. (1.2) is due to the averaging of the dipolar interaction $v(\vec{r}_1, \vec{r}_2)$ over the homogeneous probability density $1/\Omega^2$ in the volume $\Omega$ of the confinement. The coupling v of Eq. (1.3) incorporates the many-body effects in the total coupling $\bar{v}$ It involves the density fluctuation $\delta n(\vec{r}) = n(\vec{r}) - N/\Omega$ of the microscopic density

$$n(\vec{r}) = \sum_{\nu=1}^{N} \delta(\vec{r} - \vec{r}_\nu) \qquad (1.4)$$

around the average value $N/\Omega$. The bracket $\langle ... \rangle$ in Eq.(1.3) denotes for the ensemble averaging over configurations $\{\delta n(\vec{r})\}$ obeying the equilibrium distribution function. For the perfect gas, the correlation function of the density fluctuations

$$C(\vec{r}_1, \vec{r}_2) = \langle \delta n(\vec{r}_1) \delta n(\vec{r}_2) \rangle \qquad (1.5)$$

vanishes, $C(\vec{r}_1, \vec{r}_2) = 0$, hence $\bar{\upsilon}_\Delta = 0$, resulting in the expression for the average dipolar interaction of the gas of independent spins,[25] $\bar{\upsilon} = \bar{\upsilon}_0 \sim 1/\Omega$. The two mechanisms by which the environment changes the nuclear spin state of a probe $^{129}Xe$ atom, namely by the dipolar interaction between the free nuclear spin and by the dipolar interaction within the condensed spin clusters of the size about the coherence distance, are shown in Fig. 1.

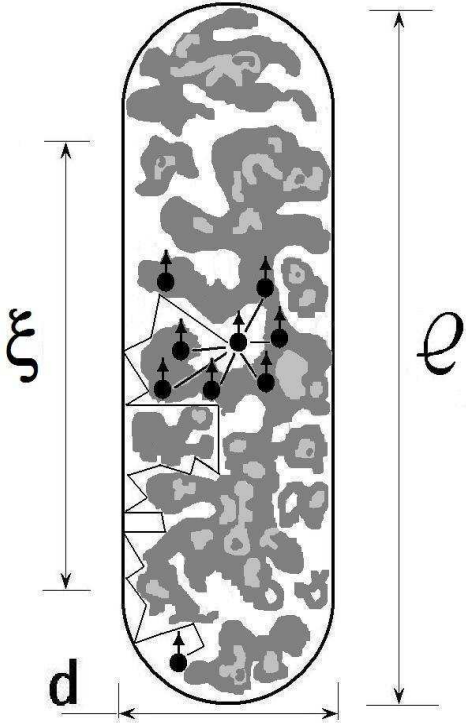

Fig. 1. Cartoon of the two mechanisms by which the environment affects the nuclear spin state of a probe (central) $^{129}Xe$ atom. First, free $^{129}Xe$ atoms not belonging to the (grey) clusters of the fluid directly alter the nuclear spin state of the probe $^{129}Xe$ atom by short range dipolar interaction after the individual atoms randomly encounter (zig-zag curves) in the tube. This interaction of the free $^{129}Xe$ atoms is superimposed on the many-body interaction (the second mechanism) when the density enhancements of the compressible fluid enhance the probability of collisions of the atoms akin the cage effect. At the CP of the fluid, the second mechanism dominates: the effective dipolar coupling between the $^{129}Xe$ atoms is mediated essentially by the dipolar interactions with the extended clusters. We denote that the origin of the effective long-range effective dipolar coupling $\bar{\upsilon}_\Delta$ at the critical point of the $Xe$ nanofluid differs from the origin of the long-range super-exchange coupling due to spin-carrying medium. Critical blob with the correlation length $\xi \approx \ell \gg d$ is almost homogeneous in any transversal cross section and inhomogeneous along the longitudinal $z$ direction.

We would now like to ask the following two questions: How can the density fluctuations affect the NMR line shape of the nanofluids? And how does to maximize the impact of the nuclear spin density fluctuations on the effective dipolar coupling? It is well documented the spectroscopic manifestations of the cooperative behavior of the particles that arises, for instance, in the critical opalescence[27] and super-radiance.[28] The principal features of the cooperative density fluctuations are embodied at the thermodynamic CP of the fluid where the density fluctuations are strong and long ranged. At the CP of the fluid, the characteristic size of the density fluctuations is extended over the size of the confinement. Consequently, the relaxation processes in the critical fluid are ruled by the density fluctuations and not by the interactions between the isolated pairs of the atoms of the fluid, so that one might expect much faster rate of the polarization transfer between the nuclear spins than the rate $\bar{\upsilon}_0$. Indeed, as is known, at the CP, the volume integrals containing the density correlation function $C(\vec{r}_1, \vec{r}_2)$ diverge in the bulk.[27]

Thus, at the first sight, the average dipolar interaction $\bar{v}_\Delta$ of Eq. (1.3) diverges providing $\bar{v} \gg \bar{v}_0$. When the finite size effects are at play, the function $C(\vec{r}_1, \vec{r}_2)$ is specified by the finite-size scaling theory[29,30] directing us to give more accurate evaluation of the volume integrals in Eq. (1.3). We show that (1) high atomic mobility and (2) the very large isothermal compressibility $n^{-1}(\partial n/\partial P)_T$ of the critical fluid within (3) the closed long nanotubes provide the correlation function $\langle \delta n(\vec{r}_1) \delta n(\vec{r}_2) \rangle \sim \ell$, see Eq. (3.30a) below, for most of the poins $\vec{r}_1$, $\vec{r}_2$ in the closed tubes of the length $\ell$. Thus, the fluctuation dipolar coupling $\bar{v}_\Delta$ of Eq. (1.3) becomes, at least, the $\ell$ times larger as compared to the coupling $\bar{v}_0 \sim 1/\ell$.

The key idea of this paper is to show that the long-range spatial coherence of the spin carrying critical nanofluid affect the nuclear spin coherence resulting in abnormally large chemical shifts $\sim \bar{v}_\Delta N$ of the nanofluid and the large intencities of the NMR signals. These properties enable supersensing in solution NMR of the critical hyperpolarized nanofluid and might be promising in molecular target recognition. Supersensing here means a large variance of the chemical shifts of the different target nuclei possessing different gyromagnetic ratio when they bind to the CNTs hyperpolarized nanofluid. The large NMR signals imply a high resolution of the critical hyperpolarized nanofluid, e.g. the Fourier of the free induction decay $\langle I_x \rangle(\omega) \sim N$ of Sec. II.C, superradiation power $P \sim \bar{v}_\Delta^2 N^3$ of Sec. II.D, and the absorbtion line shape $\chi''(\omega) \sim N$ of Sec. II.E. In providing the simultaneous effect of the spatial coherence of the fluid and the coherence of the nuclear spin states over the coherence distance $\xi \approx \ell$, a long cylindrical geometry of the critical nanofluid in the closed tube of the length $\ell$ is essential.

This idea underlies two main parts of the paper. Section II treats adiabatic regime of the nuclear spin dynamics, analysis of free induction decay, rf-superradiation, and absorption line shape of hyperpolarized nanofluid taking for granted the effective dipolar coupling constant $\bar{v}$ we develop in the Section III.

The paper is organized as follows. The many-body Liouville equation for the binary solution $^{129}Xe - ^{132}Xe$ is introduced in Section II.A. The equation describes both the quantum spin dynamics and the diffusion of $Xe$ atoms in the CNTs. Typical time $\tau_{dif}$ of the diffusive relaxation is shown to be small as compared to the characteristic time $\tau_{nmr}$ of the nuclear spin flips up and down, yielding the small adiabatic parameter

$$\varepsilon_{nmr} = \tau_{dif}/\tau_{nmr} \ll 1. \tag{1.6}$$

This is followed by the adiabatic elimination of the fast relaxing non-steady spatial states in the total density matrix. The derivation of the many-body adiabatic Liouvillean in Section II.B is based on the Zwanzig-Nakajima projection operator technique[31-33] and the perturbation theory[34,35] up to the first order in $\varepsilon_{nmr}$ resulting in the mean field spin Liouvillean with averaged dipolar coupling $\bar{v}$.

The critical thermodynamics of the fluids in long closed cylindrical tube offers a capability to achieve the finite volume-independent dipolar coupling $\bar{v}$ between all the $N$ nuclear spins of the fluid. We consider three NMR applications of abnormally large dipolar coupling $\bar{v}$ at the CP of the nanofluid. Section II.C contains a prediction of scattering of the NMR resonances in the free induction decay (FID) of $^{129}Xe$ nanofluid while approaching the CP of the nanofluid. Section II.D presents the theory of superradiation of the hyperpolarized nanofluid in two limiting cases of the low and large Larmor frequencies. This Section also shows a criterion when the radiating power in the low-field resonator exceeds the power of the Dicke's superradiation. In Section II.E we calculate the absorption line shape of the hyper-polarizeed critical nanofluid and show an abnormally large shift of the resonant frequency from the Larmor frequency.

The dipolar coupling $\bar{v}$ of Eq. (1.1) embodies the NMR spectroscopic-structural relationship due to the density correlation function $C(\vec{r}_1,\vec{r}_2)$ which enters into the expression for the coupling $\bar{v}_\Delta$ of Eq. (1.3). Derivation of an explicit expression for the function $C(\vec{r}_1,\vec{r}_2)$ at the critical point of the binary $^{129}Xe - {}^{132}Xe$ solution is outlined in Section III.D relying on the Landau-Ginzburg (LG) formulation.[27,36-38] The relevant phenomenological coupling $a$ (the inverse thermodynamic susceptibility) and $c$ (the hardness) are borrowed from the finite-size scaling theory[29,30] in Section III.C. We argue the smallness of the Ginzburg parameter

$$\varepsilon_{Gi}(T=T_c)=0 \tag{1.7}$$

at the critical temperature $T_c$ relying on the conservation of the total number of $Xe$ atoms in the sealed confinement. The existence of the two small parameters $\varepsilon_{nmr} \ll 1$ and $\varepsilon_{Gi} \ll 1$ is of importance for the calculation of the effective interaction $\bar{v}$ of Eqs. (1.1)-(1.3). Whereas the smallness $\varepsilon_{nmr} \ll 1$ for the spin degrees of freedom justifies the mean-field (MF) expression for the coupling $\bar{v}$ in terms of the correlation function $C(\vec{r}_1,\vec{r}_2)$, the smallness $\varepsilon_{Gi} \ll 1$ validates the MF (or Ornshtein-Zernike, OZ) approximati-on for the correlation function $C(\vec{r}_1,\vec{r}_2)$ of the critical solutions in the confinement. Section III.D relates the OZ correlation function $C(\vec{r}_1,\vec{r}_2)$ with the Green function $G(\vec{r}_1,\vec{r}_2)$ of the LG functional. We generalize the van Kampen's derivation[39] of the relationship between the OZ correlation function $C(\vec{r}_1,\vec{r}_2)$ and the Green function $G(\vec{r}_1,\vec{r}_2)$ to the case of an arbitrary shaped sealed confinement in order to apply the OZ relationship to the sealed pseudo-one-dimensional confinement with $\nabla G(\vec{r}_1,\vec{r}_2)=0$ on the boundary. The Green function $G(\vec{r}_1,\vec{r}_2)$ is derived in the asymptotic limit of the long tubes with the diameter $d$ far less than their length $\ell$ whereas the correlation length $\xi$ of nanofluid comes close to the length $\ell$,

$$d/\ell \ll 1, \qquad d/\xi \ll 1. \tag{1.8}$$

Given the OZ correlation function $C(\vec{r}_1,\vec{r}_2)$, we show the importance of the long cylindrical geometry of the sealed confinement for providing the abnormally large NMR line-width of the CP fluid: For the flat tubes or for nano-containers with the commensurable axes lengths, the critical spin density fluctuations of the nanofluids are irrelevant in providing an abnormally large the NMR line-width. Section III.G gives sought for dependence of the effective interaction $\bar{v}$ on the length $\ell$ of the tubes, their cross section area $A$, the critical concentration of the $^{129}Xe$ solute, and the temperature $T$. Appendices A – F comprise the auxiliary calculations of the main text, e.g. we defer the explicit calculations of the 6-fold integrals in Eqs. (1.2), (1.3) until Appendix F.

## II. ADIABATIC REGIME OF NUCLEAR SPIN DYNAMICS

### A. Total Liouvillean. Fast and slow subdynamics

Consider a solution $^{129}Xe - {}^{132}Xe$ of $N$ $^{129}Xe$ atoms (stable isotope, nuclear spin $I = \frac{1}{2}$, natural abundance 26.4 %)[40] and $M$ $(M \geq 0)$ $^{132}Xe$ atoms (stable isotope, $I = 0$, n.a. 26.9 %)[40] within the closed CNT. To orient ourselves we use the critical parameters of the bulk critical $Xe$ solution with the critical temperature $T_c \approx 16\ °C$, the total critical number density $\bar{n}_t = (N+M)/\Omega \approx 5$ atoms/nm$^3$, and the pressure about 58 atm.[41] In the temperature and pressure range of interest, the $Xe$ solution is not the ideal gas, and we want to derive at these conditions an effecttive nuclear-spin-Liouvillean $L_{eff}$ for the $^{129}Xe$ subsystem. The starting point of the derivation of the operator $L_{eff}$ is the quantum-classical Liouville equation[42] for the joint density matrix $\rho(\vec{I}^{(N)}, \vec{r}^{(N+M)}, \vec{p}^{(N+M)} t)$ of the full set of the $N$ spin operators $\vec{I}^{(N)} = \{\vec{I}_1, ..., \vec{I}_N\}$ and the $(N+M)$ coordinates and momentum vectors,

$$i\hbar \frac{d}{dt}\rho = L\rho = (L_{dz} + i\hbar L_{c\ell})\rho. \tag{2.1}$$

Here, the total Liouville operator $L$ includes the quantum $N$-nuclear spin dynamics

$$L_{dz}\rho = [H_{dz}, \rho], \tag{2.2a}$$

with the dipolar Hamiltonian[43]

$$H_{dz} = -\sum_{1 \leq n < n' \leq N} \upsilon(\vec{r}_n, \vec{r}_{n'})(3I_{nz}I_{n'z} - \vec{I}_n \vec{I}_{n'}), \tag{2.2b}$$

involving the dipolar coupling

$$\upsilon(\vec{r}_n, \vec{r}_{n'}) = \gamma^2 \hbar^2 \frac{3\cos^2\theta_{nn'} - 1}{2r_{nn'}^3}, \tag{2.2}$$

where $\gamma$ is the gyromagnetic ratio of $^{129}Xe$, the $I_{n\alpha}$ ($\alpha = x, y, z$) denotes the spin-1/2 operator[43] of $n$ th nucleus $^{129}Xe$, and $\theta_{nn'}$ refers to the polar angle between the vector $\vec{r}_{nn'} = \vec{r}_n - \vec{r}_{n'}$ and an external magnetic field $B_0$ along $z$ axis. The Liouvillean of the classical motional dynamics of the $(N+M)$ atoms $Xe$ reads[32,33]

$$L_{c\ell} = -\sum_{1 \leq n \leq N+M}\left(\frac{\vec{p}_n}{m_n}\frac{\partial}{\partial \vec{r}_n} + \vec{f}_n \frac{\partial}{\partial \vec{p}_n}\right), \tag{2.3}$$

with the force $\vec{f}_n = -\partial U/\partial \vec{r}_n$ on the potential energy landscape $U$ of all the atoms,

$$U(\vec{r}^{(N+M)}) = \sum_{1 \leq n < n' \leq N+M} u(\vec{r}_n, \vec{r}_{n'}). \tag{2.4}$$

The density matrix of Eq. (2.1) is diagonal with respect to all the classical coordinates and momentum vectors. The atoms $^{129}Xe$ and $^{132}Xe$ are regarded to be interacting via the common Lennard-Jones (LJ) potential $u(\vec{r}_n, \vec{r}_{n'})$ so that the binary mixture $^{129}Xe - {}^{132}Xe$ is thought of as composed of the two thermodynamically indistinguishable fluids. It follows that the binary solution $^{129}Xe - {}^{132}Xe$, from the thermodynamic point of view, is a single component $Xe$ fluid. This notion simplifies the derivation the isothermal susceptibility of the critical solute $^{129}Xe$ in terms of the susceptibility of the entire critical solution, see Section III.B.

At the first glance, the introduction of the coordinates and momentum of the spin-carrying nuclei in Eq. (2.1) makes the description of the solution NMR signals more elaborate, as compared to the case of motionless spins. Nevertheless, it is well known the stochastic motion of spin carrying atoms greatly simplifies the spin dynamics by narrowing the dipolar broadening and producing high resolution NMR spectra of liquids.[25,43-47] Rather than removing the dipolar broadening, our strategy in the paper is, quite the contrary, to enhance the dipolar broadening as far as possible even if we break the MF condition of Eq. (1.6). In this way we appeal to the adiabatic theory. Its usefulness lies in the capability to separate out the dynamics of the external (coordinates and momentum) variables from the dynamics of the internal (spin) variables in the total density matrix $\rho$. Really, the total Liouvillean $L$ for the confined $Xe$ fluid is characterized by the interplay of three distant frequency scales.

Irreversible diffusive subdynamics of the total density matrix $\rho$ in $\vec{p}^{(N+M)}$-subspace reaches the Maxwell distribution on the time scale,[48]

$$\tau_v = \lambda / \langle v \rangle \approx 10^{-11} \text{ s}, \quad (2.5)$$

where the mean free path $\lambda = \left( \bar{n}_t \pi a_0^2 \right)^{-1} \approx 1 \text{ nm}$ is estimated for the total critical number density $\bar{n}_t \approx 5 \text{ nm}^{-3}$, the $a_0 \approx 0.436 \text{ nm}$ stands for the interatomic separation at the LJ well depth,[41] $\varepsilon / k_B \approx 250 \text{ K}$, and the $\langle v \rangle = \sqrt{3 k_B T_c / m} \approx 10^2$ m/s is the thermal velocity of the $Xe$ atoms at the thermodynamic equilibrium, with $k_B$ being the Boltzmann's constant.

Irreversible relaxation in the $\vec{r}^{(N+M)}$-subspace to the Boltzmann distribution proceeds over the diffusional timescale

$$\tau_{\text{dif}} \approx \ell^2 / D_{\text{self}} \approx \ell^2 / (\lambda \langle v \rangle) \approx 10^{-5} \text{ s}, \quad (2.6)$$

which refers to the time it takes for a nucleus to spread a distance of typical nanotube length $\ell \approx 1 \ \mu\text{m}$. In deriving the estimation (2.6), we make use the well-known relation for self diffusion coefficient $D \approx \lambda \langle v \rangle \approx 10^{-7} \text{ m}^2/s$ in the bulk. For the CNTs, it has been observed the super-diffusion[8] which is by three orders of magnitude faster than the above value $D \approx 10^{-7} \text{ m}^2/s$ due to the slip flow on smooth walls of the CNTs, thus, the estimate $\tau_{\text{dif}} \approx 10^{-5} \text{ s}$ in Eq. (2.6) can be diminished to the value of the order $\tau_{\text{dif}} \approx 10^{-8} \text{ s}$. The relaxation to the steady state Boltzmann distribution function in $\vec{r}^{(N+M)}$-subspace goes by the Smoluchowski Liouvillean[49]

$$L_{c\ell} = D \sum_{1 \leq n \leq N+M} \frac{\partial}{\partial \vec{r}_n} \left( \frac{\partial}{\partial \vec{r}_n} - \frac{1}{k_B T} \vec{f}_n \right), \tag{2.7}$$

which replaces the Liouvillean of Eq. (2.3) over the time[47] $t \gg mD/(k_B T) = \lambda/\langle v \rangle = \tau_v$. From Eqs. (2.5), (2.6) it follows that, over the time scale $\tau_{dif}$ ($\gg \tau_v$), the subdynamics of the classical part of the total density matrix $\rho$ reaches the Maxwell-Boltzmann distribution

$$\rho_{cl} = \exp(-E/k_B T) / \iint d\vec{r}^{(N+M)} d\vec{p}^{(N+M)} \exp(-E/k_B T), \tag{2.8}$$

with the total energy being $E = \sum_{v=1}^{N+M} \vec{p}_v^2/(2m_v) + U(\vec{r}^{(N+M)})$ and the normalization of the distribution $\rho_{cl}$ to unity.

Flipping up and down of the nuclear spins occur on the characteristic time scale

$$\tau_{nmr} = \left( \gamma^2 \hbar / \langle \ell_{nn} \rangle^3 \right)^{-1} = \left( \chi \, \gamma^2 \hbar / a_0^3 \right)^{-1}, \tag{2.9}$$

that is well separated from the time scales $\tau_v$ of Eq. (2.5) and $\tau_{dif}$ of Eq. (2.6). In the relation (2.9), the length $\langle \ell_{nn} \rangle$ is the most probable distance between nearest neighbor spin-carrying atoms $^{129}Xe$ in the solution $^{129}Xe - ^{132}Xe$, accordingly, the $\chi$ represents the volume fraction occupied by the solute $^{129}Xe$ in the solution,

$$\chi = a_0^3 / \langle \ell_{nn} \rangle^3 = a_0^3 N/\Omega = a_0^3 \bar{n}, \tag{2.10}$$

where $\bar{n} = N/\Omega$ is the number density of the $^{129}Xe$ atoms in the solution $^{129}Xe - ^{132}Xe$ within the closed container of the volume $\Omega$. For $^{129}Xe$, $\gamma^2 \hbar / a_0^3 \approx 700$ rad/s [50] hence the time scale $\tau_{nmr}$ of Eq. (2.9) is far much larger than the time scales $\tau_v$ of Eq. (2.5) and $\tau_{dif}$ of Eq. (2.6), especially for the diluted solution of $^{129}Xe$ with $\chi \ll 1$. The smallness of the adiabatic parameter $\varepsilon_{nmr} = \tau_{dif}/\tau_{nmr} \ll 1$ of Eq. (1.6) underlies the separation of the fast relaxing variables $\vec{p}^{(N+M)}$ and $\vec{r}^{(N+M)}$ from the slow nuclear spin variables $\vec{I}^{(N)}$ in the total density matrix $\rho$ while deriving an effective spin Liouvillean from the total Liouvillian $L$ of Eq. (2.1).

### B. Adiabatic spin Liouvillean

Adiabatic elimination of fast relaxing degrees of freedom is the powerful approach in reducing the number of degrees of freedom in dynamical many-body systems having far separated characteristic time scales belonging to the different degrees of freedom. The adiabatic theory has been successful in developing the theory of superradiance,[51] the spectra of liquids,[35] Brownian motion,[52] chemical master equations.[49,53] The similarity between these elementary processes description and the description of nuclear spin dynamics within nanocontainers permits invoking the adiabatic methodology to get insight into the NMR of fluctuating nanofluids. In order to pass on to the description of the slow nuclear spin variables $\vec{I}^{(N)}$ one should to integrate out the fast relaxing variables $\vec{r}^{(N+M)}$ and $\vec{p}^{(N+M)}$ in the total density matrix. We, first, calculate the $x$-polarization

$$\langle I_x \rangle (t) = \iint d\vec{r}^{(N+M)} d\vec{p}^{(N+M)} \, \text{tr}\left\{ I_x \rho\left(\vec{I}^{(N)}, \vec{r}^{(N+M)}, \vec{p}^{(N+M)}, t\right) \right\} \tag{2.11}$$

of the collective spin $I_x = \sum_{\nu=1}^{N} I_{\nu,x}$ (in units of $\hbar$). We prove in this Section that the $\langle I_x \rangle (t)$ of Eq. (2.11) can be cast into a much more simple form

$$\langle I_x \rangle (t) = \text{tr}\left\{ I_x \sigma(\vec{I}, t) \right\} \tag{2.12}$$

involving the reduced density matrix $\sigma(t) = \sigma(\vec{I}, t)$ independent of the coordinates and momentum of the nuclear spins. The physical content of the dependence of the state $\sigma$ on the collective spin operators $I_a = \sum_{\nu=1}^{N} I_{\nu,a}$, $a = x, y, z$, lies in the fast motion of spin-carrying nuclei. Due to intermixing of all the $N$ nuclei in the closed confinement, the average magnetic field produced by $\nu$-th spin at a point $\vec{r}$ is independent of the spin's number $\nu$. Thus, on the coarse grained time $\tau_{\text{dif}}$, the whole spin ensemble behaves, for the purposes of bringing the average dipolar magnetic field, as a single unit. It would be satisfactory to separate out the fast coordinate variables of the nuclei from their slow spin variables in the way

$$\rho\left(\vec{I}^{(N)}, \vec{r}^{(N+M)}, \vec{p}^{(N+M)}, t\right) \xrightarrow{t \gg \tau_{\text{dif}}} \rho_{c\ell}\left(\vec{r}^{(N+M)}, \vec{p}^{(N+M)}\right) \cdot \sigma(\vec{I}, t), \tag{2.13}$$

where the distribution $\rho_{c\ell}$ is given by Eq. (2.8). Herein we prove the relation (2.13) relying on the Zwanzig-Nakajiama projection technique[31-33], and as by-product, we derive the effective spin-Liouvillean of the evolution of the reduced density matrix $\sigma$. One defines by $P$,

$$P\{...\} = \rho_{c\ell} \iint d\vec{r}^{(N+M)} d\vec{p}^{(N+M)} \{...\}, \quad P^2 = P, \tag{2.14}$$

the projector[34] onto the steady state $\rho_{c\ell}$ of the classical transport operator $L_{c\ell}$ of Eq. (2.7), with

$$L_{c\ell} \rho_{c\ell} = 0. \tag{2.15}$$

Action of the operator $P$ upon an arbitrary density matrix $\rho\left(\vec{I}^{(N)}, \vec{r}^{(N+M)}, \vec{p}^{(N+M)}, t\right)$ reads

$$P\rho = \rho_{c\ell} \cdot \sigma(\vec{I}^{(N)}, t), \tag{2.16}$$

with

$$\sigma(\vec{I}^{(N)}, t) = \iint d\vec{r}^{(N+M)} d\vec{p}^{(N+M)} \rho, \tag{2.17}$$

so that Eq. (2.13) indicates that

$$\rho\left(\vec{I}^{(N)},\vec{r}^{(N+M)},\vec{p}^{(N+M)},t\right)\xrightarrow{t\gg\tau_{dif}}P\rho=\rho_{c\ell}\cdot\sigma, \qquad (2.18)$$

Before writing down the equation of motion for the relevant density matrix $P\rho$,[31] we introduce the identities

$$L_{c\ell}P=0, \qquad (2.19)$$

and

$$PL_{c\ell}=0. \qquad (2.20)$$

The property of Eq. (2.19) is the straightforward consequence of that $\rho_{c\ell}$ is the right eigenvector of the operator $L_{c\ell}$ with zero eigenvalue. The property of Eq. (2.20) shows that projector $P$ is also the left eigenvector of the operator $L_{c\ell}$ with zero eigenvalue resulting from the conservation with time of the total number of $Xe$ atoms in the closed confinement. Define finally the projector $Q$ onto the non-steady states of the transport operator $L_{c\ell}$,

$$Q=1-P, \qquad Q^2=Q. \qquad (2.21)$$

With projectors $P$ of Eq. (2.14) and $Q$ of Eq. (2.21), we immediately draw the exact equation[31] for the relevant density matrix $(\hbar=1)$,

$$i\frac{d}{dt}\left(P\rho(t)\right)=PLe^{-itQL}\left(Q\rho(0)\right)+PL\left(P\rho(t)\right)-i\int_0^t ds PLe^{-isQL}QL\left(P\rho(t-s)\right). \qquad (2.22)$$

Now we prove in the limit that $\varepsilon_{nmr}\ll 1$ and for times $t\gg\tau_{dif}$, the Eq. (2.22) reduces to the closed equation for the relevant density matrix $P\rho$,

$$i\frac{d}{dt}\left(P\rho(t)\right)=\left(PL_{dz}P\right)\left(P\rho(t)\right). \qquad (2.23)$$

To prove Eq. (2.23), we need some preparations. First, we estimate the inhomogeneous term of Eq. (2.22) making use of the smallness of the parameter $\varepsilon_{nmr}\ll 1$ which can be rewritten in terms of the norms of the operators involved as follows

$$\varepsilon_{nmr}=\|L_{dz}\|/\|L_{c\ell}\|\ll 1. \qquad (2.24)$$

By Eqs. (2.24), (2.21), and (2.20) we estimate

$$QL=QL_{dz}+iL_{c\ell}\approx iL_{c\ell}. \qquad (2.25)$$

It follows that the non-steady state $Q\rho(0)$ diminish with time as $(t\gg\tau_{dif})$,

$$e^{-itQL}\left(Q\rho(0)\right)\approx e^{tL_{c\ell}}\left(Q\rho(0)\right)\approx e^{-t/\tau_{dif}}\left(Q\rho(0)\right)\to 0. \qquad (2.26)$$

The lesson of this is that the initial conditions are irrelevant for the effective dynamics over time $t \gg \tau_{dif}$ and the term $e^{-itQL}(Q\rho(0))$ of Eq. (2.22) can be dropped out. The integral term of Eq. (2.22) contains the slowly varying density matrix $P\rho(t)$ and, by Eq. (2.26), the short-time memory operator $e^{-itQL}Q$. The resultant equation can be rewritten[34,35] as the first order equation for $P\rho(t)$ at elapsed time $t$,

$$i\frac{d}{dt}(P\rho(t)) = \left(PL_{dz}P - PL_{dz}Q(iL_{c\ell})^{-1}QL_{dz} + O(\varepsilon_{nmr}^2)\right)(P\rho(t)), \tag{2.27}$$

The leading term of the order $\varepsilon_{nmr}^0$ is provided by the first term in the bracket in the r.-hand side of Eq. (2.27), hence the integral term in Eq. (2.22) can be discarded when treating the effective Liouvillean to the lowest nontrivial order in $\varepsilon_{nmr}$. We denote in passing that all (nonadiabatic) terms of the perturbation expansion of the effective Liouvillean in $\varepsilon_{nmr}$ are given by the Bloch perturbation theory.[54] To establish the governing equation for the spin density matrix $\sigma$ we following[34,35] substitute Eq. (2.16) in Eq. (2.23) and make use the projective condition (2.14), yielding

$$i\rho_{c\ell}\frac{d}{dt}\sigma = PL_{dz}P\cdot\sigma = \rho_{c\ell}\left(\iint d\vec{r}^{(N+M)}d\vec{p}^{(N+M)}\rho_{c\ell}L_{dz}\right)\sigma. \tag{2.28}$$

Cancelling the common factor $\rho_{c\ell}$ in Eq. (2.28), we arrive at sought for equation for the spin density matrix $\sigma$

$$i\frac{d}{dt}\sigma = \bar{L}_{dz}\sigma, \tag{2.29a}$$

with adiabatic spin Liouvillean

$$\bar{L}_{dz} = \iint d\vec{r}^{(N+M)}d\vec{p}^{(N+M)}\rho_{c\ell}L_{dz}. \tag{2.29b}$$

Referring to the Hamiltonian description, Eqs. (2.29) are equivalent to the reversible spin dynamics

$$i\frac{d}{dt}\sigma = \left[\bar{H}_{dz},\sigma\right] \tag{2.30a}$$

with the adiabatic spin Hamiltonian

$$\bar{H}_{dz} = \iint d\vec{r}^{(N+M)}d\vec{p}^{(N+M)}\rho_{c\ell}H_{dz}. \tag{2.30b}$$

Operator $\bar{H}_{dz}$ comes also if one starts, from the very beginning, with the Liouville-von Neumann equation

$$i\frac{d}{dt}\rho = \left[H_{dz}\left(\vec{I}^{(N)},\vec{r}^{(N+M)}(t)\right),\rho\right] \tag{2.31}$$

with the dipolar Hamiltonian $H_{dz}$ depending implicitly on time through the coordinates $\vec{r}^{(N+M)}(t)$ obeying the classical equations of motion. Temporary averaging of Eq. (2.31) over the rapid stochastic coordinates with foregoing invoking of the ergodic hypothesis yields the adiabatic Hamiltonian $\bar{H}_{dz}$.[25] The ergodic condition is valid at the steady state when each configuration $\{\vec{r}^{(N+M)}, \vec{p}^{(N+M)}\}$ encounters in the Gibbs ensemble with the probability $\rho_{c\ell}$ of Eq. (2.8), hence the operator $\bar{H}_{dz}$ of Eq. (2.30b). The spin Hamiltonian $\bar{H}_{dz}$ of Eq. (2.30) can be rephrased in terms of the collective spin operators $I_a$ and $\vec{I}^2 = I_x^2 + I_y^2 + I_z^2$ as follows

$$\bar{H}_{dz} = -\tfrac{1}{2}\bar{\upsilon} \sum_{1 \leq n \neq n' \leq N} \left(3I_{nz}I_{n'z} - \vec{I}_n \vec{I}_{n'}\right) = -\tfrac{1}{2}\bar{\upsilon}\left(3I_z^2 - \vec{I}^2\right), \tag{2.32}$$

with a unique dipolar coupling constant

$$\bar{\upsilon} = \int d\vec{r}_n d\vec{r}_{n'}\, \rho_{c\ell}^{(2)}(\vec{r}_n, \vec{r}_{n'}) \upsilon(\vec{r}_n, \vec{r}_{n'}) \tag{2.33}$$

for all the pairs $n \neq n' = 1,...,N$ of the nuclear spins, where we defined the reduced distribution

$$\rho_{c\ell}^{(2)}(\vec{r}_1, \vec{r}_2) = \int d\vec{r}_3...d\vec{r}_{N+M} \rho_{c\ell}\left(\vec{r}^{(N+M)}\right). \tag{2.34}$$

In obtaining Eqs. (2.33), (2.34) we carried out integrals over the momentum space in the expression $\bar{H}_{dz}$ of Eq. (2.30) and, hence, retain only the coordinate dependence of the distribution function $\rho_{c\ell}$ of Eq. (2.8), then, while carrying out the remaining configuration integral in the expression $\bar{H}_{dz}$ of Eq. (2.30), we integrate only the functions $\upsilon(\vec{r}_n, \vec{r}_{n'})$ since each spin operator $I_{n\alpha}$ associated with the $n$-th nucleus does not depend explicitly on its coordinate $\vec{r}_n$, and, finally, we take advantage of the symmetry of the function $\rho_{c\ell}\left(\vec{r}^{(N+M)}\right)$ with respect to the permutations of all the identical spin carrying nuclei. The coupling $\bar{\upsilon}$ of Eq. (2.33) is the Gibbs ensemble averaged or the motionally averaged dipolar interaction $\upsilon(\vec{r}_n, \vec{r}_{n'})$. The qualitative picture behind this definition is that the coupling $\bar{\upsilon}$ for any pair of the spins $(n, n')$ 'senses' the whole distribution of the nuclei $n$ and $n'$ over the domain of their rapid motion, but not the instant positions $(\vec{r}_n, \vec{r}_{n'})$ of the nuclei $n$ and $n'$. Eqs. (2.33) and (2.34) have already been given[55] as a generalization of the ergodic arguments,[25] i.e. without invoking the adiabatic theory presented in this Section. In closing this Section we prove the relation of Eq. (2.12). To this end, we substitute Eq. (2.18) in Eq. (2.11), where the density matrix $\sigma$ is governed by Eq. (2.30) with the Hamiltonian $\bar{H}_{dz}$ of Eq. (2.32) and the distribution $\rho_{c\ell}$ of Eq. (2.8). In doing so, we straightforwardly get the $x$ polarization $\langle I_x \rangle(t)$ of Eq. (2.12).

### C. FID of the hyperpolarized nanofluid

We want to calculate the FID of the hyperpolarized spin nanofluid in order to confirm the well known assertion[56] that this state of spin carrying nanofluid is more favorable as compared to the nearly unpolarized nanofluid at high temperatures owing to much stronger FID signal that can be detected from the small number of spins of the confined nanofluid. Here is the FID and its Fourier transform $\langle I_x \rangle(\omega)$ calculated in a wide temperature range far and near the critical point of the hyperpolarized CNTs nanonofluid. We predict the scattering of resonances of the function $\langle I_x \rangle(\omega)$ when approaching the CP of the nanofluid and describe quasi-continuous Gaussian shape[25] far from the CP.

Let the polarization of all the $N$ nuclear spins of the nanofluid be along the $z$-axis,

$$\sigma = |0\rangle\langle 0|, \qquad (2.38)$$

Where $|0\rangle = \prod_{n=1}^{N} |0\rangle_n$ stands for the down spin state with $|0\rangle_n = \begin{pmatrix} 0 \\ 1 \end{pmatrix}_n$. The density matrix $\sigma$ is invariant with time since the state $|0\rangle$ is the eigenstate $|I = N/2, m = -N/2\rangle$ of the operators $I_z$ and $\bar{H}_{dz}$ of Eq. (2.32) in the angular momentum representation. Let $t = 0$ be the time moment of $\pi/2$ hard pulse about the $Y$ axis of the laboratory frame so that the density matrix after the pulse becomes

$$\sigma(+0) = e^{i\frac{\pi}{2}I_Y} |0\rangle\langle 0| e^{-i\frac{\pi}{2}I_Y}. \qquad (2.39)$$

The system with the density matrix $\sigma(t) = e^{-iH_0 t}\sigma(+0)e^{iH_0 t}$ subjected to the effective Hamiltonian $H_0 = -\omega_0 I_z + \bar{H}_{dz}$ is allowed to decay resulting in the x-polarization $\langle I_x \rangle(t)$ of Eq. (2.12) of the form, see Eq. (A8),

$$\langle I_x \rangle(t) = \cos(\omega_0 t)\langle 0| e^{-i\frac{\pi}{2}I_Y} e^{i\bar{H}_{dz}t} I_x e^{-i\bar{H}_{dz}t} e^{i\frac{\pi}{2}I_Y} |0\rangle \qquad (2.40)$$

$$= (N/2)\cos(\omega_0 t)\left(\cos\left(\tfrac{3}{2}\bar{\upsilon}t\right)\right)^{N-1}. \qquad (2.41)$$

The FID of Eq. (2.41) is $2k_B T/(\hbar\omega_0) \approx 10^6$ times stronger than the FID[55]

$$\langle I_x \rangle(t) = \frac{\hbar\omega_0}{k_B T}\cos(\omega_0 t)\,\mathrm{tr}\left\{I_x e^{-i\bar{H}_{dz}t} I_x e^{i\bar{H}_{dz}t}\right\}\Big/\mathrm{tr}\{1\}$$

$$= \frac{\hbar\omega_0}{k_B T}\frac{N}{4}\cos(\omega_0 t)\left(\cos\left(\tfrac{3}{2}\bar{\upsilon}t\right)\right)^{N-1} \qquad (2.42)$$

originated from the high temperature state[43] $\sigma(t=0) = \left(1 - \frac{\hbar\omega_0}{k_B T}I_z\right)\Big/\mathrm{tr}\{1\}$ at temperature $T = 300K$ and Larmor frequency[50] $|\omega_0| = |\gamma_{^{129}Xe}|B_0 = 11.78$ MHz at field $B_0 = 1$ T.

We substitute in Eq. (2.41) the identity $\cos\varphi = (e^{i\varphi} + e^{-i\varphi})/2$, $\varphi = \tfrac{3}{2}\bar{\upsilon}t$, and use the binomial expansion in order to recast the FID in the way

$$\langle I_x \rangle(t) = (N/2)\cos(\omega_0 t) \sum_{n=-(N-1)/2}^{(N-1)/2} A_n \, e^{3in\bar{\upsilon}t} \tag{2.43}$$

with coefficients

$$A_n = \frac{1}{2^{N-1}} \binom{N-1}{\frac{N-1}{2}+n} \approx \frac{1}{\sqrt{\pi N/2}} e^{-n^2/(N/2)} \,. \tag{2.44}$$

Fourier transforming Eq. (2.43) and applying $\int_{-\infty}^{\infty} dt\, e^{i\Omega t} = 2\pi\delta(\Omega)$ give the function

$$\langle I_x \rangle(\omega) = (N/2)\pi \sum_{n=-(N-1)/2}^{(N-1)/2} A_n \, \delta(\omega - \omega_n) \tag{2.45}$$

in the form of the lattice of the $\delta$-peaks at equidistant frequencies $\omega_n$ centered at the Larmor frequency,

$$\omega_n = \omega_0 + 3\bar{\upsilon}n, \; n = -(N-1)/2,\ldots,(N-1)/2 \,, \tag{2.46}$$

and the Gaussian envelope of the intensities $A_n$ of Eq. (2.44). The set of the $\delta$-peaks merges into the quasi-continuous Gaussian function with the spacing $\omega_{n+1} - \omega_n = 3\bar{\upsilon} \to 0$ between the peaks when the effective coupling $\bar{\upsilon} \to 0$. Increase in the effective coupling $\bar{\upsilon}$ results in the broadening of the Gaussian shape $\langle I_x \rangle(\omega)$ of Eq. (2.45) so that it breaks up into the sum of well identified resonant $\delta$-peaks. We will show in Section III.G, see Eqs. (3.50), that the situation when $\bar{\upsilon} \approx \gamma^2 \hbar/\Omega \to 0$ and when $\bar{\upsilon}$ is a finite quantity can be observed far and near the CP of the $^{129}Xe$ nanofluid, respectively. This property underlies the effect of the scattering of the resonances of the hyperpolarized nanofluid as we go to the CP, see Fig. 2.

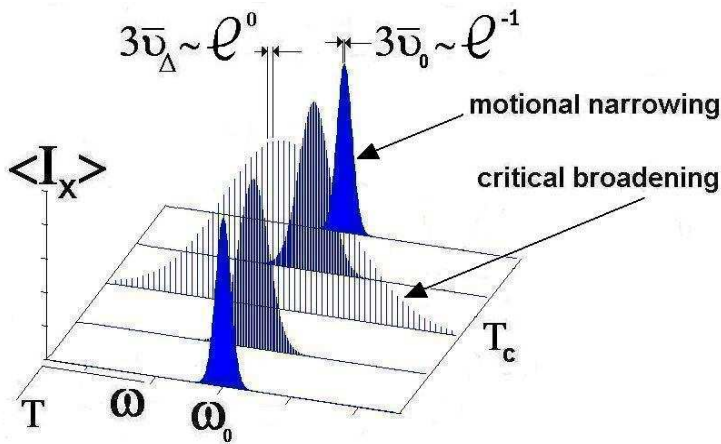

Fig. 2. Representation of Eq. (2.45). Fluctuation induced scattering of the NMR resonances of the hyperpolarized $^{129}Xe$ nanofluid when the temperature of the nanofluid passes across the critical temperature $T_c \approx 16\; °C$.

Far from the CP of the CNTs nanofluid (hereafter CNT's large axis is assumed to parallel to the field $B_0$), the dipolar coupling $\bar{\upsilon} = 2\pi\gamma^2\hbar/\Omega$, see Eq. (3.50) of Sec. III.G, thus the resonant line has the width $3\bar{v}N = 2\pi \cdot 3\gamma^2 \hbar \bar{n}$ $(< 150 \text{ Hz})$, with $\bar{n} = N/\Omega$ referring to the density of $^{129}Xe$ atoms in the CNTs nanofluid.

In the vicinity of the CP of the nanofluid, the width $3\bar{v}N$ is the extensive quantity $\sim N$ since the effective coupling $\bar{v}$ is a finite volume-independent quantity, $\bar{v} = \text{const}(\Omega) \approx 10 \text{ Hz}$, see Eq. (3.50). We can view this behavior of the FID as the issue of a specific interplay between the quantum mechanical nuclear spin states and the classical nuclear density fluctuations at the critical state of the fluid. The discrete structure of the function $\langle I_x \rangle(\omega)$ is provided here not by a special rf-irradiation of the fluid but due to the finite residual dipolar coupling $\bar{v}$ between all the nuclei of the critical fluid.

### D. Super-emf voltage

Hyperpolarized CNTs nanofluid at the critical point possess the coherences both over the spin states and the spatial degrees of freedom simultaneously and as *per se* is a unique object with a maximized NMR characteristics such as the intensity and frequency of the NMR signals. We want to discuss in this Section a spontaneous decay of the polarization $\langle I_x \rangle(t)$ and to show a radiofrequency superradiance of the critical nanofluids at zero and low Larmor frequencies as the counterpart of the Dicke's superradiance.[28] Let the hyperpolarized nanofluid is prepared at the low Larmor frequency either by the dynamical polarization[57] or by the adiabatic demagnetization technique[43,57] over a time-span before the moment $t = 0$. For example, the adiabatic demagnetization eliminates the external magnetic field $B_0$ with the rate $\left| B_0^{-1} dB_0/dt \right| \ll \bar{v}$ while retaining the secular effective Hamiltonian $\bar{H}_{dz}$ of Eq. (2.32) and without loss of already preconditioned hyperpolarization. This is because the hyperpolarized states $|1\rangle$ and $|0\rangle$ are the eigenstates of the operators $I_z$ and $\bar{H}_{dz}$ simultaneously, with the eigenstates being $|I = N/2, m = \pm N/2\rangle$ in the angular momentum basis, respectively. After applying the $(\pi/2)_Y$ strong pulse at $t = 0$, the hyperpolarized nanofluid is specified by the $x$-polarization $\langle I_x \rangle(t)$ of Eq. (2.41) with the low Larmor frequency $|\omega_0| = |\gamma_{^{129}Xe}| B_0$. By Eq. (2.41), for $|\omega_0| \ll \bar{v}\sqrt{N}$ or $\omega_0 = 0$, the induced electromotive force (emf) at the ends of the pick-up coil enclosing the nanofluid is proportional to the value

$$d\langle I_x \rangle(t)/dt = -\tfrac{3}{4}\bar{v}N(N-1)\cos(\omega_0 t)\sin\left(\tfrac{3}{2}\bar{v}t\right)\left(\cos\left(\tfrac{3}{2}\bar{v}t\right)\right)^{N-2}. \quad (2.47)$$

In the vicinity of the CP of the nanofluid, the effective dipolar coupling has a finite value $\bar{v} = \text{const}(\Omega) \approx 10 \text{ Hz}$, see Eq. (3.50), hence the factor $\bar{v}N$ in Eq. (2.47) is an extensive quantity, $\sim N$, and one should expect an abnormally large derivative $d\langle I_x \rangle(t)/dt$. Indeed, careful calculation gives maximal $d\langle I_x \rangle(t)/dt = -\tfrac{3}{4}\bar{v}N^{3/2}/\sqrt{e}$. To prove this, we represent the periodic function $\left(\cos\left(\tfrac{3}{2}\bar{v}t\right)\right)^{N-2}$ in the form of repeating Gaussian peaks,

$$\left(\cos\left(\tfrac{3}{2}\overline{\upsilon}t\right)\right)^{N-2} = \sum_{k=0}^{\infty} e^{-\left(t-\frac{2\pi}{3\overline{\upsilon}}k\right)^2 / \tau^2} \tag{2.48}$$

for even large $(N-2)$ and short time scale

$$\tau = 2\sqrt{2}/\left(3\overline{\upsilon}\sqrt{N}\right), \tag{2.49}$$

which is conditioned by $\tfrac{3}{2}\overline{\upsilon}\tau \ll \pi$. We treat the first Gaussian peak with $k=0$ or near $t=0$ so that Eq. (2.47) becomes

$$d\langle I_x\rangle(t)/dt = -\tfrac{3}{4}\overline{\upsilon}N^2 \cos(\omega_0 t) e^{-t^2/\tau^2} \sin\left(\tfrac{3}{2}\overline{\upsilon}t\right). \tag{2.50}$$

The function $e^{-t^2/\tau^2}\sin\left(\tfrac{3}{2}\overline{\upsilon}t\right)$ in the right-hand side of Eq. (2.51) growths from $0$ at $t=0$ reaches the maximum value

$$\max\left\{e^{-t^2/\tau^2}\sin\left(\tfrac{3}{2}\overline{\upsilon}t\right)\right\} = \left(\sqrt{e}\right)^{-1}\sin\left(1/\sqrt{N}\right) \approx 1/\sqrt{eN} \tag{2.51}$$

at the moment $t_* = \tau/\sqrt{2} = 2/\left(3\overline{\upsilon}\sqrt{N}\right)$ and then drops to zero for $t > t_*$. By Eq. (2.51), the derivative $d\langle I_x\rangle(t)/dt$ of Eq. (2.50) at the burst is

$$\max\left\{d\langle I_x\rangle(t)/dt\right\} = -\tfrac{3}{4}\overline{\upsilon}N^{3/2}/\sqrt{e} \tag{2.52}$$

at the moment $t_*$. The same conclusion comes from Eq. (2.49) and Eq. (2.41) with $\omega_0 = 0$ if one analyses the maximum value of the function $\max\left\{d\langle I_x\rangle(t)/dt\right\}$ as follows

$$\max\left\{\langle I_x\rangle(t)\right\}/\tau = -(N/2)/\tau = -\tfrac{3}{4}\overline{\upsilon}N^{3/2}/\sqrt{2}, \tag{2.53}$$

which is largely concerned with the careful calculations of Eq. (2.52). By Eqs. (2.48), (2.52) the sought function $d\langle I_x\rangle(t)/dt$ of Eq. (2.47) is the periodic function with the amplitude $\tfrac{3}{4}\overline{\upsilon}N^{3/2}/\sqrt{e}$ at the bursts separated by the time interval $2\pi/(3\overline{\upsilon})$.

The derivation of Eq. (2.53) looks very much like the qualitative explanation of the rate of the spontaneous superradiation[28] where the role of the rate $d\langle I_x\rangle(t)/dt$ is played by the quantity $d\langle I_z\rangle(t)/dt$ which is the rate of changing the spin population difference. This quantity is related with the radiation power $P_D = -\hbar\omega d\langle I_z\rangle(t)/dt$ of the resonant quanta $\hbar\omega$.[58] At the bursts, the radiation power $P_D$ scales as the square of the number of radiating atoms, $P_D = \hbar\omega N/\tau_D = \hbar\omega^2 N^2$ where the Dicke's time scale of the superradiation is $\tau_D = 1/(\omega N)$.[28] Comparing the power $P_D$ with Eq. (2.53) shows that if the Dicke's time scale $\tau_D = 1/(\omega N)$ denotes the characteristic time of the longitudinal relaxation then the time scale $\tau = 2\sqrt{2}/(3\bar{v}\sqrt{N})$ of Eq. (2.49) denotes the characteristic time of the transverse relaxation, the so called $T_2$. Superposition principle[28,58] claims that the maximal emf voltage is proportional to $N|\omega_0|$.[43] When $\omega_0 = 0$, the depolarization $\langle I_x \rangle(t)$ proceeds exactly by Eq. (2.47), or by its corollary Eq. (2.52), yielding the scaling $\sim \bar{v} N^{3/2}$ rather than the scaling $\sim N \cdot 0$. For $|\omega_0| \ll \bar{v}\sqrt{N}$, the spin ensemble precesses in the inherent dipolar field $\sim \bar{v} N^{1/2} |\gamma|^{-1}$ rather than in the external magnetic field which is vanishingly small: the polarization $\langle I_x \rangle(t)$ can never be at rest all the time causing a non-zero spontaneous emf voltage in the pick-up coil.

By Eq. (2.52), we can calculate the induced emf voltage[59]

$$V(t) = -\mu_0 \left( N_{(\ell)\text{coil}}/\ell \right) \gamma \hbar \left( d\langle I_x \rangle(t)/dt \right), \tag{2.54}$$

where the factor $\mathcal{B}_{1x} = \mu_0 \left( N_{(\ell)\text{coil}}/\ell \right)$ T/A denotes the magnetic field within the solenoid with $\left( N_{(\ell)\text{coil}}/\ell \right)$ number of turns of the wire per meter in the coil and unit current, the $\mu_0 = 4\pi \cdot 10^{-7}$ T·m/A refers to the magnetic permeability of free space, and $|\gamma|\hbar = 7.4 \cdot 10^{-27}$ A·m² stands for the magnetic dipole moment of $^{129}Xe$ nucleus[50] (in SI units). Putting into the expression for $V(t)$ of Eq. (2.54) the value $\left( N_{(\ell)\text{coil}}/\ell \right) \approx 10^6$ turns/m corresponding to the coil with the microscale wire[60] and the expression for $\max\{d\langle I_x \rangle(t)/dt\}$ of Eq. (2.52) with $\bar{v} \approx 62.8$ rad/s and $N \approx 10^{14}$ number of $^{129}Xe$ atoms corresponding to the critical density $\bar{n} \approx 5$ nm$^{-3}$ within the cylindrical microdroplet of the volume $\Omega \approx (30\mu\text{m})^3$ leads to $\max\{V(t)\} \approx 400$ $\mu$V at the bursts. The same order-of-magnitude voltage can be obtained if $10^7$ critical cylindrical microdroplets each with $N \approx 10^9$ number of $^{129}Xe$ nuclei are assembled in parallel within the pick-up coil. The calculated emf voltage is hardly greater than few $\mu$V fluctuations caused by the thermal noise in the coil of wire at room temperature, thus suggesting a favorable verification.

Far from the CP of the nanofluid, the effective coupling reads $\bar{v} = 2\pi \gamma^2 \hbar/\Omega$, see Eq. (3.50) of Sec. III.G below, so that the factor $\bar{v} N^2$ in Eq. (2.50) scales with $N$ as follows $\bar{v} N^2 = 2\pi \gamma^2 \hbar \bar{n} N$. Invoking Eq. (2.51), we have the rate $d\langle I_x \rangle(t)/dt$ of Eq. (2.50) at the burst

$$\max\{d\langle I_x \rangle(t)/dt\} = -\tfrac{3\pi}{2} \bar{v} N^{1/2}/\sqrt{e}. \tag{2.55}$$

The induced emf voltage $V(t)$ of Eq. (2.54) generates the current $\mathfrak{I}(t) = V(t)/R$ dissipated with the power $P = V^2/R$ in the pick-up coil with the resistance $R$. By Eqs. (2.52) and (2.55), the dissipated power at the bursts is

$$P = \left(\frac{3}{4\sqrt{e}} \mu_0 \frac{N_{(\ell)\text{coil}}}{\ell} \gamma \hbar\right)^2 \frac{1}{R} \cdot \begin{cases} \bar{v}^2 \cdot N^3 & \text{- at the CP} \\ \left(2\pi\gamma^2 \hbar \bar{n}\right)^2 \cdot N & \text{- far from the CP} \end{cases} \qquad (2.56)$$

for the coherent rate of the energy dissipation ($\xi_\ell \approx \ell$, top line) and for the incoherent loss of the energy ($\xi_\ell \ll \ell$, bottom line of Eq. (2.56)). For $\bar{v} \approx 62.8$ rad/s, $\gamma^2 \hbar \approx 58.5$ rad/s nm$^3$,[50] and $\bar{n} \approx 5$ nm$^{-3}$ we have the dimensionless ratio $\bar{v}^2 / \left(2\pi\gamma^2 \hbar \bar{n}\right)^2 \approx 10^{-3}$ in Eq. (2.56).

In deriving Eqs. (2.52), (2.55), and (2.56), we have treated the spontaneous radiation of the nanofluids in the zero-field or in low-field resonators, i.e. $\omega_0 = 0$ or $|\omega_0| \ll \bar{v}\sqrt{N}$ cases study. In the opposite limiting case $|\omega_0| \gg \bar{v}\sqrt{N}$, the hyperpolarization $\langle I_x \rangle(t)$ of Eq. (2.41) has the rate

$$d\langle I_x\rangle(t)/dt = -(N/2)\omega_0 \sin(\omega_0 t)\left(\cos\left(\tfrac{3}{2}\bar{v}t\right)\right)^{N-1}, \qquad (2.57)$$

so that $\max\{d\langle I_x\rangle(t)/dt\} = -(N/2)|\omega_0|$ at the burst.[43] It is this rate of the depolarization along side with the rate $d\langle I_x\rangle(t)/dt$ of Eq. (2.52) that allows one to discriminate between the low frequency $\left(|\omega_0| \ll \bar{v}\sqrt{N}\right)$ and high frequency $\left(|\omega_0| \gg \bar{v}\sqrt{N}\right)$ rate $d\langle I_x\rangle(t)/dt$. In Eq. (2.52) the role of the Zeeman field $\omega_0 |\gamma|^{-1}$ is played by the dipolar field $\sim \bar{v} N^{1/2} |\gamma|^{-1}$. This dipolar field is $\sqrt{N}$ times smaller than the dipolar field $\bar{v} N |\gamma|^{-1}$ of the fully porarized nuclei indicating that the dipolar field $\sim \bar{v} N^{1/2} |\gamma|^{-1}$ can be understood as originating from the dispersion of the $N$ random spins tumbling in the $x$-$y$ plane.

To keep distinction between the low frequency depolarization of the CP nanofluid, see Eq. (2.52), and the high-frequency depolarization of Eq. (2.57), we calculate the ratio of the amplitudes of the emf voltages,

$$\frac{\max\{V_D\}}{\max\{V\}} = \left(2\sqrt{e}/3\right)\left(\frac{|\omega_0|}{\bar{v}\sqrt{N}}\right), \qquad (2.58)$$

where the amplitude in the high frequency case is $\max\{V_D(t)\} = \mu_0 \left(N_{(\ell)\text{coil}}/\ell\right)(N/2)|\omega_0|$.[43] Analogously, the Dicke's power[28]

$$P_D = V_D^2 R^{-1} \sim \left(d\langle I_x\rangle(t)/dt\right)^2 = \omega_0^2 N^2 / 4 \qquad (2.59)$$

admits the comparison with the power for the CP nanofluids without the resonator, see Eq. (2.56),

$$\frac{P_D}{P} = \left(2\sqrt{e}/3\right)^2 \left(\frac{\omega_0}{\bar{v}\sqrt{N}}\right)^2. \qquad (2.60)$$

By Eq. (2.60), for $N \approx 10^{14}$ number of $^{129}Xe$ nuclei of the critical cylindrical nanofluid with $\bar{v} \approx 10$ Hz and $|\omega_0| \approx 12$ MHz ($B_0 = 1$ T) we have $P_D \approx 10^{-2} P$.

In searching for the low field superradiation which exceeds the Dicke's superradiation, it is more favorable to deal with the spin systems with a large gyromagnetic ratio $\gamma$ since $\bar{\upsilon} \sim \gamma^2$. Calculation the ratio of Eq. (2.60) for $N \approx 10^{14}$ number of hyperpolarized unpaired electron spins of the critical cylindrical microfluid with dipolar coupling $\bar{\upsilon} \approx \left(g\mu_B/\gamma_{^{129}Xe}\right)^2 \cdot 10$ Hz $\approx 10^7$ Hz and the resonator specified by the frequency $\omega_0 \approx 10^{10}$ Hz (microwave length $\lambda \approx 3$ cm) results in $P_D \approx 10^{-8} P$, with $g = 2$ and $\mu_B$ being the Bohr magneton.

Preceding calculations of the spontaneous depolarization $\langle I_x \rangle(t)$ of Eq. (2.41) neglect the feedback of the induced emf current $\Im$ on the depolarization.[61] This neglect of the feedback is justified when the wire of the pick-up coil has so large resistance $R$ that the induced emf current $\Im$ is small, hence, the induced magnetic field $\hat{b}_x = \mu_0 \left(N_{(\ell)\text{coil}}/\ell\right) \Im(t)$ of the solenoid is negligible as compared with the characteristic dipolar field $\bar{\upsilon} N^{1/2} |\gamma|^{-1}$ of the nanofluid. For the above example of $N \approx 10^{14}$ spins of the critical hyperpolarized $^{129}Xe$ nanofluid with $\bar{\upsilon} \approx 10$ Hz, the coil should have the resistance $R \gg 10^{-2}$ Ohm for the spontaneous depolarization to be valid.

**E. Absorption line shape of hyperpolarized naninofluid**

We prove in this Section that the absorption line shape $\chi''(\omega)$ of the hyperpolarized $^{129}Xe$ nanofluid in the nuclear spin state $|0\rangle = \prod_{n=1}^{N} \begin{pmatrix} 0 \\ 1 \end{pmatrix}_n = |I = N/2, m = -N/2\rangle$ has the form

$$\chi''(\omega) = \gamma^2 \frac{\pi N}{4} \left(\delta(\omega + \omega_*) - \delta(\omega - \omega_*)\right) \tag{2.61a}$$

with the resonant frequency being

$$\omega_* = \omega_0 - \tfrac{3}{2} \bar{\upsilon}(N-1). \tag{2.61b}$$

For $^{129}Xe$ nuclei in the state $|0\rangle$, the resonant frequency $\omega_*$ is negative definite because the Larmor frequency $\omega_0 = \gamma_{^{129}Xe} \cdot B_0 < 0$,[50] and the coupling $\bar{\upsilon} \sim \gamma^2 > 0$, see Eq. (3.50). It follows that the function $\chi''(\omega)$ has the only positively valued $\delta$-peak in the physical domain $\omega \geq 0$ at the frequency $\omega = -\omega_* > 0$. We also notice that the line shape $\chi''(\omega)$ of Eqs. (2.61) quite differs from the Fourier of the FID $\langle I_x \rangle(\omega)$ of Eq. (2.45). This is contrasted with the case of almost unpolarized (high temperature) nanofluid where the absorption line shape $\chi''(\omega)$ has the same frequency dependence[43,44] as the function $\langle I_x \rangle(\omega)$.

By treating the Liouville-von Neumann Eq. (2.30a) with the Hamiltonian $(\hbar = 1)$

$$H = -\omega_0 I_z + \bar{H}_{dz} - \gamma B_x I_x \cos(\omega t) \tag{2.62}$$

and the initial density matrix in the remote past

$$\sigma(t = -\infty) = |0\rangle\langle 0|, \tag{2.63}$$

we appeal to the standard recipe of the linear response technique to the first order in the magnetic field $B_x$. The result is this[43,62]

$$\sigma(t) = |0\rangle\langle 0| + i\gamma B_x \int_0^\infty dt' \cos(\omega(t-t'))[I_x(t'), |0\rangle\langle 0|]. \tag{2.64}$$

Here, we have introduced the operator $I_x(t') = e^{-iH_0 t'} I_x e^{iH_0 t'}$ in the interaction picture and the Hamiltonian $H_0 = -\omega_0 I_z + \bar{H}_{dz}$. In deriving Eq. (2.64) we have used the commutator $[|0\rangle\langle 0|, H_0] = 0$, hence $e^{iH_0 t'}|0\rangle\langle 0|e^{-iH_0 t'} = |0\rangle\langle 0|$. The generic form of the $x$-polarization after the perturbation of the eigenstate $|0\rangle$ with the weak magnetic field $B_x \cos(\omega t)$ reads[43]

$$\gamma\langle I_x\rangle(t) = \gamma \operatorname{tr}\{I_x \sigma(t)\} = B_x(\chi'(\omega)\cos(\omega t) + \chi''(\omega)\sin(\omega t)).$$

On substituting here the density matrix $\sigma(t)$ of Eq. (2.64) one has the line shape $\chi''(\omega)$ in terms of the correlation function $\langle 0|I_x I_x(t')|0\rangle$,

$$\chi''(\omega) = -2\gamma^2 \int_0^\infty dt' \sin(\omega t') \operatorname{Im}\{\langle 0|I_x I_x(t')|0\rangle\}. \tag{2.65}$$

Here we make use, firstly, the identities $\operatorname{tr}\{I_x I_x(t')|0\rangle\langle 0|\} = (\operatorname{tr}\{|0\rangle\langle 0|I_x(t')I_x\})^+ = \langle 0|I_x I_x(t')|0\rangle$, secondly, the zero value of the initial $x$-polarization $\operatorname{tr}\{I_x|0\rangle\langle 0|\} = 0$, and finally, the cyclic reordering under the trace. Straightforward calculation of Eq. (2.65) leaves us with the function $\chi''(\omega)$ of Eqs. (2.61), see Appendix C. The meaning of the resonance frequency $\omega_*$ of Eq. (2.61b) is easy to understand. A subsystem of $(N-1)$ nuclear spins in the state $|0\rangle_n = |m_n = -1/2\rangle$ with $\gamma < 0$ generates the up directed dipolar magnetic field resulting in the increase of the total magnetic, see Fig. 3(a). The enhancement of the total magnetic field causes the increase of the resonant frequency of the remaining probe spin from the value $|\omega_0| = |\gamma_{129_{Xe}}|B_0$ up to $|\omega_*|$ of Eq. (2.61b). The positive value of the function $\chi''(\omega)$ at the resonant frequency $\omega = -\omega_*$ means that the system in the state $|0\rangle$ (this is the lowest Zeeman energy state of all the $N$ nuclear spins in the external magnetic field $B_0$) absorbs rf-quanta at the frequency $\omega = -\omega_*$. Comparing the function $\chi''(\omega)$ of Eq. (2.61a) with the function $\langle I_x\rangle(\omega)$ of Eq. (2.45) shows that the two types of the excitations of the $N$ nuclear ensemble give two different responses.

The line shape $\chi''(\omega)$ of Eq. (2.61) is obtained under the weak continuous perturbation of the ensemble of 'strongly' coupled spins responding at the single resonant frequency $\omega = -\omega_*$ of Eq. (2.61b). This line shape $\chi''(\omega)$ distinguishes from the FID $\langle I_x \rangle(\omega)$ of Eqs. (2.45), (2.46). The function $\langle I_x \rangle(\omega)$ exhibits all the $N$ resonances, with the maximal resonant frequency coinciding with the resonant frequency $\omega = -\omega_*$ of Eq. (2.61b).

Fig. 3. Illustrations to Eq. (2.61b) and Eq. (2.67b). (a) The magnetic moments $\mu_z = \hbar \gamma m_n$ of $^{129}Xe$ nuclear spins in the state $|0\rangle_n = |m_n = -1/2\rangle$ with the gyromagnetic ratio $\gamma < 0$ point upward (wide grey arrow $\Uparrow$) and augment the up directed external magnetic field $B_0$. For any number $N$ of the nuclear spins, the state $|0\rangle$ absorbs the resonant quanta. (b) Magnetic moments $\mu_z$ of $^{129}Xe$ nuclear spins in the

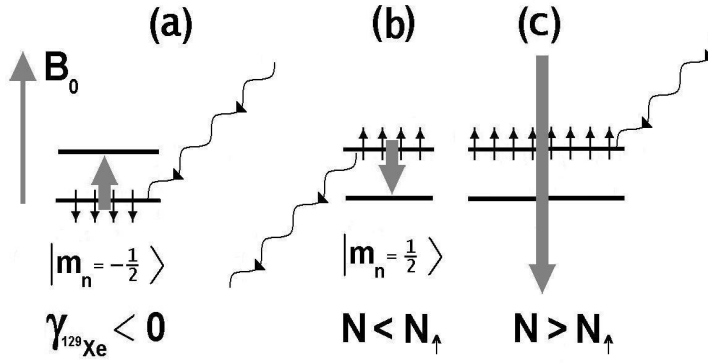

state $|1\rangle_n = |m_n = 1/2\rangle$ are antiparallel ($\Downarrow$) to the magnetic field $B_0$. This state is allowed to energy loss by emission of the resonant quanta. (c) If the number $N$ of the nuclear spins exceeds the value $N_\downarrow$ of Eq. (2.68), the inherent dipolar magnetic field of the nuclear spins inverts the total magnetic field and the state $|1\rangle$ becomes absorbing.

It is interesting to compare the line shape $\chi''(\omega)$ of Eqs. (2.61) calculated for the initial condition $\sigma(t = -\infty) = |0\rangle\langle 0|$ with the line shape for the initial condition $\sigma(t = -\infty) = |1\rangle\langle 1|$ where $|1\rangle = \prod_{n=1}^{N} \begin{pmatrix} 1 \\ 0 \end{pmatrix}_n$. The eigenstate $|1\rangle$ of $^{129}Xe$ fluid is the highest Zeeman energy state of all the $N$ up directed nuclear spins $\begin{pmatrix} 1 \\ 0 \end{pmatrix}_n = |m_n = 1/2\rangle$ having $\gamma < 0$, see Fig. 3(b). For the eigenstate $|1\rangle$, the absorption line shape reads

$$\chi''(\omega) = \gamma^2 \frac{\pi N}{4} \left( \delta(\omega - \omega_{**}) - \delta(\omega + \omega_{**}) \right) \qquad (2.67a)$$

with the resonant frequency

$$\omega_{**} = \omega_0 + \tfrac{3}{2} \overline{v}(N-1). \qquad (2.67b)$$

A subsystem of $(N-1)$ spins in the state $|1\rangle_n = |m_n = 1/2\rangle$ generates an inherent dipolar magnetic field antiparallel to the external magnetic field $B_0$. This dipolar magnetic field weakens the total magnetic field and causing the decrease of the resonant frequency of the remaining probe spin from the value $|\omega_0| = |\gamma_{^{129}Xe}| B_0$ to $|\omega_{**}|$ of Eq. (2.67b). If the number $N$ is less than the $N_\downarrow$ specified by the relationship $|\gamma_{^{129}Xe}| B_0 = \frac{3}{2} \bar{\upsilon}(N_\downarrow - 1)$ or

$$N_\downarrow = 1 + \tfrac{2}{3} |\gamma_{^{129}Xe}| B_0 / \bar{\upsilon}_\Delta , \qquad (2.68)$$

the $N$ nuclear fluid emit the resonant quanta at the frequency $\omega = -\omega_{**} > 0$ so that the line shape $\chi''(\omega = -\omega_{**}) = -\gamma^2 \dfrac{\pi N}{4} \delta(\omega + \omega_{**}) < 0$. Otherwise, if $N > N_\downarrow$, the inherent dipolar magnetic field of the fluid inverts the total magnetic field so that the state $|1\rangle$ becomes energetically lowest in the total magnetic field and, hence, the state $|1\rangle$ absorbs rf-quanta at the frequency $\omega = \omega_{**} > 0$, yielding $\chi''(\omega = \omega_{**}) = \gamma^2 \dfrac{\pi N}{4} \delta(\omega - \omega_{**}) > 0$, see Figs. 3(b), 3(c). The effect of the inversion of the total magnetic field is due to the dipolar interactions only: the effect is missing if the spins are effectively isolated, $\bar{\upsilon} = 0$, or the nanofluid is far from the CP when $\bar{\upsilon} \sim \Omega^{-1} \to 0$ in the long tubes. For the case of the critical hyperpolarized cylindrical $^{129}Xe$ nanofluid with $\bar{\upsilon} \approx 10$ Hz, the critical number of the nuclear spins providing the inversion of the total magnetic field is $N_\downarrow \approx 0.78 \cdot 10^6$ for $B_0 = 1$ T.

The resonant frequency $\omega_*$ is temperature responsive near the CP of the CNTs nanofluid. This property lies at the root of the sensitivity of the NMR cw- spectrum of the critical hyperpolarized nanofluid. The detuning $\omega_0 - \omega_* = \frac{3}{2} \bar{\upsilon}(N-1)$ of Eq. (2.62) is an extensive quantity $\sim N$ at the CP where $\bar{\upsilon} = \bar{\upsilon}_\Delta \approx 10$ Hz is the finite volume-independent coupling. The deturning $\omega_0 - \omega_*$ is the intensive quantity $3\pi\gamma^2 \hbar \bar{n}$ ($< 75$ Hz) far from the CP where $\bar{\upsilon} = 2\pi\gamma^2 \hbar / \Omega$, see Eq. (3.50) of Sec. III.G.

To appreciate the strength of $z$-directed dipolar field caused by the critical hyperpolarized cylindrical $^{129}Xe$ nanofluid, we set $N = 10^9$ and $\bar{\upsilon} \approx 10$ Hz in the formula for the dipolar frequency shift $\frac{3}{2} \bar{\upsilon} N \approx 15$ GHz. This value is 50 times larger than $^{129}Xe$ Larmor frequency $\omega_0 \approx 280$ MHz of 23.5 T present-day NMR magnet.[63]

### III. CRITICAL THERMODYNAMICS OF CLOSED CNTs NANOFLUID

#### A. Dipolar coupling – correlation function relationship

The NMR signals of the nanofluids require a prior knowledge of the effective dipolar coupling $\bar{\upsilon}$ of Eq. (2.33). We therefore proceed to that problem by recasting, first, the effective coupling $\bar{\upsilon}$ in the way that it becomes related to the density correlation function of $^{129}Xe$ fluid,

$$\bar{\upsilon} = \frac{1}{N(N-1)} \iint_\Omega d^3\vec{r}_1 d^3\vec{r}_2 \ \upsilon(\vec{r}_1,\vec{r}_2) \langle n(\vec{r}_1) n(\vec{r}_2) \rangle, \qquad (3.1)$$

where the relation[27,64] $\rho_{c\ell}^{(2)}(\vec{r}_1,\vec{r}_2) = \langle n(\vec{r}_1) n(\vec{r}_2) \rangle / (N(N-1))$, is used for $\vec{r}_1 \neq \vec{r}_2$, see Eq. (C7). This modification of the expression for the coupling $\bar{\upsilon}$ of Eq. (2.33) is more appropriate for an explicit calculation of the coupling $\bar{\upsilon}$. It is the dependence on the density correlation function that is the characteristic of the many body systems and that is useful in a phenomenological description of the pairing correlations in the strongly interacting systems.[27,37,64] The coupling $\bar{\upsilon}$ of Eq. (3.1) is nothing but the ensemble average of the total dipolar energy per pair of equivalent spin-carrying $^{129}Xe$ atoms. In Eq. (3.1), we single out the contribution $\bar{\upsilon}_0$ coming from the independent atoms making use of Eq. (C9), yielding

$$\bar{\upsilon} = \bar{\upsilon}_0 + \bar{\upsilon}_\Delta, \qquad (3.2)$$

with couplings $\bar{\upsilon}_0$ and $\bar{\upsilon}_\Delta$ obeying Eq. (1.2) and Eq. (1.3), respectively. The decomposition of the total coupling $\bar{\upsilon}$ of Eq. (3.2) into the two terms of Eqs. (1.2), (1.3) standing for the perfect gas and the correlatons of the atoms is the central result of this Section. This decomposition makes it evident that all the dependence of the effective coupling $\bar{\upsilon}$ on the temperature $T$ and the concentration of $Xe$ atoms are embodied in the coupling $\bar{\upsilon}_\Delta$ of Eq. (1.3) whereas the coupling $\bar{\upsilon}_0$ of Eq. (1.2) depends only on the 'external' operational parameters: the volume and shape of the nanocontainer. Fluctuation coupling $\bar{\upsilon}_\Delta$ also depends on the geometry of the container, e.g. for cubes with the sides $\ell$ and for disks with the diameter $\ell$, the increase of the macroscopic length $\ell$ results in vanishing of the microscopic couplings $\bar{\upsilon}_\Delta$ of the critical fluid, see Eqs. (3.30) below. Meanwile, for closed long tubes of the length $\ell$, the residual coupling $\bar{\upsilon}_\Delta$ is the finite $\ell$-independent quantity, see Eq. (3.50), with coupling $\bar{\upsilon}_0$ being negligible.

Far from the CP, the correlation function $C(\vec{r}_1,\vec{r}_2)$ of Eq. (1.5) is of short-ranged resulting in the cutoff of the integration domain in the integrals of Eq. (1.3) and the property $\bar{\upsilon}_\Delta \ll \bar{\upsilon}_0$. In the vicinity of the CP, the correlation function $C(\vec{r}_1,\vec{r}_2)$ has the long-ranged behavior so that the integration domains in Eq. (1.3) and Eq. (1.2) coincide. We will show in Sec. III.D that at the CP, the correlation function scales with the nanotube's length $\ell$ as

$$C(\vec{r}_1,\vec{r}_2) \approx (\chi_T/\Omega) \cdot \mathrm{const}(\vec{r}_1,\vec{r}_2) \sim \ell \qquad (3.3)$$

since the MF isothermal susceptibility scales as $\chi_T \sim \ell^2$.[29,30] Eq. (3.3) is a succinct form in which we suppress the dependence on the cross section area $A$ of the CNTs containing $N = \bar{n} A \ell$ $^{129}Xe$ atoms. Given the scaling of the correlation function of Eq. (3.3), the probability $C(\vec{r}_1,\vec{r}_2)/N^2 \sim \ell/(\bar{n}A\ell)^2 \sim \ell^{-1}$ in Eq. (1.3) becomes larger than the probability $1/\Omega^2 \sim \ell^{-2}$ in Eq. (1.2) when we are concerned with the CP nanofluids within the long CNTs. It follows that the dependence of the term $v(\vec{r}_1,\vec{r}_2)C(\vec{r}_1,\vec{r}_2)$ in Eq. (1.3) not only on the local points $\vec{r}_1$ and $\vec{r}_2$ but also on the strong long-ranged correlation function of Eq.(3.3) involving the mesoscopic length $\ell$ results in the relationship $\overline{v}_\Delta/\overline{v}_0 \sim \ell$. This relationship is in the heart of the fluctuation-based NMR sensing of the CP nanofluids.

### B. Landau-Ginzburg functional for nanofluids

Correlation function $C(\vec{r}_1,\vec{r}_2)$ of Eq. (1.5) calls for the formalism based on the density functional framework. For this purpose, the Landau-Ginzburg (LG) formulation is the conventional one for treating the two phase region close to the liquid-gas phase transition. In applying the LG formulation, central in our calculations is the restriction with the MF approximation,[27,36-38] in part, because the MF is the engineering approximation which is capable of explaining the second order phase transition, second, because the MF scaling exponent $\eta = 0$ only slightly differs from the renormalization-group value $\eta = 0.03$ for bulk,[38] thirdly, because of the vanishing of the Ginzburg parameter for the sealed fluids, see Sec. III.F. The LG formulation for the correlation function $C(\vec{r}_1,\vec{r}_2)$ of continuous medium rests on the free energy functional expansion in the density fluctuation $\delta n(\vec{r})$ of $^{129}Xe$ atoms around the spatially homogeneous critical value $\overline{n} = N/\Omega$,

$$F(\{\overline{n} + \delta n(\vec{r})\}) = \Omega f(\overline{n},T) + \delta F, \tag{3.4a}$$

where

$$\delta F = \int_\Omega d^3\vec{r}\, \delta f(\delta n(\vec{r})), \tag{3.4b}$$

with

$$\delta f = \left(\tfrac{1}{2}a(T)(\delta n(\vec{r}))^2 + \tfrac{1}{4}b(T)(\delta n(\vec{r}))^4 + \tfrac{1}{2}c(T)(\vec{\nabla}\delta n(\vec{r}))^2\right). \tag{3.4c}$$

Few comments are in order in deriving the free energy of Eqs. (3.4). We single out the fluctuations $\delta n(\vec{r})$ of $^{129}Xe$ atoms assuming the thermodynamically indistinguishable components $^{132}Xe$ and $^{129}Xe$; eventually the inclusion of the fluctuations of $^{132}Xe$ atoms results in the renormalization of the phenomenological parameters in Eqs. (3.4c). Then, the neglecting of the surface contribution to the $F$ seems to be a good approximation for $Xe$ atoms in the CNTs due to the weak interactions between the closed shell $Xe$ atoms and the walls of the CNTs giving rise to the slipping of the simple fluids in the CNTs.[8] Finally, we write out the inverse susceptibility $a(T)$, the hardness $c(T)$, and the parameter $b(T)$ in the succinct form keeping only the dependence on temperature and leaving off the arguments denoting the average densities of the $^{129}Xe$ and $^{132}Xe$ nanofluids. In order to derive the parameters $a(T)$ and $c(T)$ for the $^{129}Xe$ nanofluid within the closed CNTs we recapitulate the LG formulation for the bulk followed by modifying the bulk values $a(T)$ and $c(T)$ invoking the Fisher's finite-size scaling.[29,30]

With shorthand

$$\eta(\vec{r}) = \delta n(\vec{r}), \tag{3.5}$$

the minimum of the $\delta F$ of Eqs. (3.4) occurs at the spatially homogeneous density fluctuation

$$\eta_* = \begin{cases} \pm\sqrt{-a/b}, & \text{for } a<0 \ (T<T_c) \\ 0, & \text{for } a \geq 0 \ (T \geq T_c) \end{cases} \tag{3.6}$$

where the function $a(T)$ near the critical temperature $T_c$ reads[27]

$$a(T) = \alpha\tau, \quad \alpha>0, \quad \tau=(T-T_c)/T_c, \tag{3.7}$$

see Fig. 4.

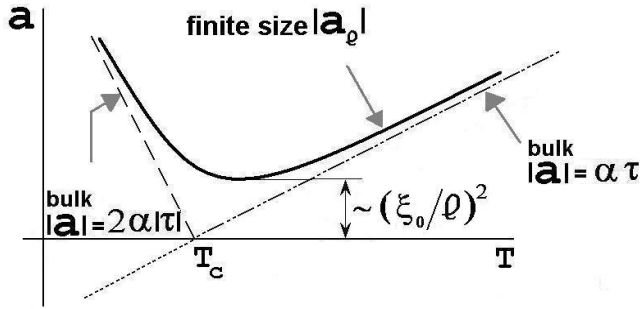

Fig. 4. The inverse susceptibilities. Dot dashed line: bulk $a(T) = \alpha\tau$ of Eq. (3.7); long dashed: bulk $|a|$ of Eq. (3.18); bold: finite size $|a_\ell|$ of Eqs. (3.23).

Expansion of the spatially homogeneous part of the free energy density of Eq. (3.4c) around the most probable value up to the quadratic terms[27]

$$\delta f(\eta) = \begin{cases} -a^2/(4b) - a(\eta-\eta_*)^2, & \text{for } a<0 \ (T<T_c) \\ \frac{1}{2}a\eta^2, & \text{for } a \geq 0 \ (T \geq T_c) \end{cases}, \tag{3.8}$$

gives the inverse susceptibility of the most probable free energy density $\delta f(\eta_*)$ of Eq. (3.8), see Fig. 5,

$$|a| = d^2(\delta f)/d\eta^2\Big|_{\eta=\eta_*} = \alpha|\tau| \cdot \vartheta(\tau), \tag{3.9}$$

with coefficient

$$\vartheta(\tau) = \begin{cases} 2, & \text{for } \tau<0 \\ 1, & \text{for } \tau>0 \end{cases}. \tag{3.10}$$

In averaging over configurations $\{\eta(\vec{r})\}$ with probability density $e^{-\delta F/k_B T}/\int D\{\eta\}e^{-\delta F/k_B T}$, the additive constant $a^2/(4b)$ in Eq. (3.8) is irrelevant so that we are left with[27]

$$\delta F = \frac{1}{2}\int_\Omega d^3\vec{r}\,\eta(\vec{r})\left(|a|-c\Delta_{\vec{r}}\right)\eta(\vec{r}). \tag{3.11}$$

In fact, the $\eta(\vec{r})$ in Eq. (3.11) should be replaced with $(\eta(\vec{r})-\eta_*)$ in describing the cases $T<T_c$, $T\geq T_c$. We omit this replacement as irrelevant in calculating the correlation function $\langle\eta(\vec{r}_1)\eta(\vec{r}_2)\rangle$. In deriving Eq. (3.11) we have used also the zero flux condition $\nabla\eta(\vec{r})=0$ on the boundary of closed confinement and the Gauss's formula for the volume integral.

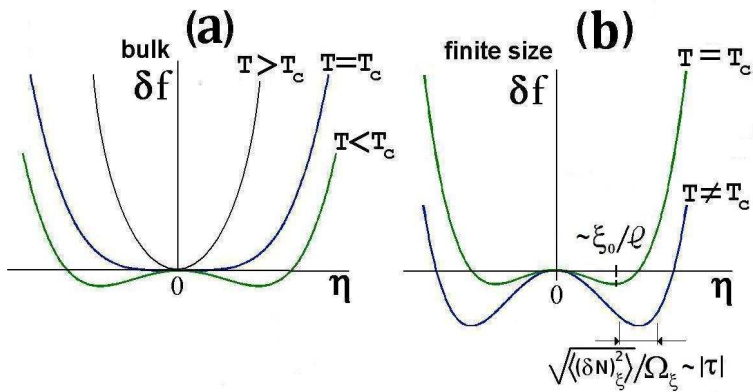

Fig. 5. Free energy density $\delta f(\eta)$ for the bulk (a) and confined (b) fluids. (a) The basins of the functions $\delta f(\eta)$ are the regions where the quadratic expansion of Eq. (3.8) validates. At the CP in the bulk, $T=T_c$, the interactions of the fast atoms with the clusters are barrierless resulting in the large scale density fluctuations. (b) Finite-size free energy density $\delta f$ has the double-well LG potential for all the temperatures. For the confined fluids, there is no temperature when the densities of the liquid and gas phases coincide, thus, the interphase boundary remains and the isothermal compressibility is large but finite. The equilibrium density fluctuation $\eta_*\sim\xi_0/\ell$ and the fluctuation of the total number of $^{129}Xe$ atoms $\sqrt{\langle(\delta N)_\xi^2\rangle}$ in the volume $\Omega_\xi = A\xi$ refer to Eq. (3.44).

For the time being we want to relate the parameter $a$, belonging to the $^{129}Xe$ component, with the pressure and concentration of the entire mixture $^{129}Xe - ^{132}Xe$. We have

$$a = \partial^2 f/\partial\bar{n}^2 = (\partial\mu_n/\partial\bar{n}), \tag{3.12}$$

with the derivatives being performed at fixed $\{T,\Omega\}$. Chemical potential $\mu_n$ of the $^{129}Xe$ component is derived using the MF total free energy $F$, namely

$$\mu_n = (\partial F/\partial N) = (\partial(f\Omega)/\partial N) = (\partial f/\partial\bar{n}). \tag{3.13}$$

To get an explicit expression for the parameter $a$ use is made of the Gibbs thermodynamic relationship,[65]

$$a = \partial \mu_n / \partial \bar{n} = (\partial \mu_n / \partial P)(\partial P / \partial \bar{n}) = \frac{1}{\bar{n}}(\partial P / \partial \bar{n}), \qquad (3.14)$$

where the relation $(\partial P / \partial \mu_n) = \bar{n}$ is due to the differential of the total pressure[27] $dP = sdT + \bar{n}d\mu_n + \bar{m}d\mu_m$ with $s = S/\Omega$ standing for the specific entropy, $\bar{m} = M/\Omega$ and $\mu_m$ are the density and chemical potential of the $^{132}Xe$ atoms, respectively, and $\bar{n} = N/\Omega$ refers to the concentration of $^{129}Xe$ atoms. Thermodynamic indistinguishability of the two $Xe$ isotopes in the $^{129}Xe - ^{132}Xe$ solution allows us to relate the differential $d\bar{n}$ in Eq. (3.14) with the differential $d\bar{n}_t = d\bar{n} + d\bar{m}$ of the total density of $Xe$ atoms putting $d\bar{n} = xd\bar{n}_t$ and keeping the fraction

$$x = N/(N + M) = \bar{n}/\bar{n}_t \qquad (3.15)$$

of $^{129}Xe$ atoms fixed, $x \leq 1$. By introducing the fraction $x$, the parameter $a$ of Eq. (3.14) becomes

$$a = \frac{1}{x^2 \bar{n}_t}\left(\frac{\partial P}{\partial \bar{n}_t}\right). \qquad (3.16)$$

Eq. (3.16) is sought for relation between the parameter $a$ of the $^{129}Xe$ subsystem with the total pressure $P$ and the total concentration $\bar{n}_t$ of $Xe$ atoms in the solution $^{129}Xe - ^{132}Xe$. Since the solution, from the thermodynamic point of view, is thought of as single component system, we immediately write out the derivative of Eqs. (3.16), (3.14) in the MF approximation as[36]

$$(\partial P / \partial \bar{n}_t) = k_B T_c (\xi_0 / \xi)^2 = k_B T_c |\tau|, \qquad (3.17)$$

with $\xi = \xi_0 / \sqrt{|\tau|}$ standing for the bulk MF correlation length near the CP and $\xi_0 \approx 1$ nm refers to the correlation length far from the CP at $|\tau| \approx 1$.[39,66] Hereafter we use the convention that in deriving the MF scaling relations we put scaling amplitudes equal to unity offering the dependence only on physical parameters involved. Eqs. (3.17), (3.16), (3.9) yield the LG inverse susceptibility at the CP,

$$|a| = \left(\frac{\xi_0}{\xi}\right)^2 \frac{k_B T_c}{x^2 \bar{n}_t} \vartheta(\tau), \qquad (3.18)$$

or equivalently as $|a| = |\tau| k_B T_c / (x^2 \bar{n}_t) \vartheta(\tau)$, see Fig. 4. Thermodynamic parameters $|a|$ of Eq. (3.18) and $c$ of Eq. (3.4c) specify, by the dimension arguments, the correlation length

$$\xi = \sqrt{c/|a|} \qquad (3.19)$$

giving the rigidity

$$c = \xi_0^2 \frac{k_B T_c}{x^2 \bar{n}_t} \vartheta(\tau), \qquad (3.20)$$

which is the finite value at the CP in accord with the density functional theory.[27,37]

### C. Confinement parameters $|a_\ell|$ and $c_\ell$ for CNTs nanofluid

Given the bulk parameters $|a|$ and $c$ of of Eq. (3.18) and Eq. (3.20), respectively, we use the finite size scaling[29,30] in order to derive the parameters $|a_\ell|$ and $c_\ell$ for the critical nanofluids within the CNTs. The CNTs are regarded as the hollow cylinders long enough in one dimension and confined in the cross sections of an arbitrary geometry. We exclude strictly 1-D systems with short range interactions because of the lack of the second order phase transitions in these systems.[27] Minimal diameter $d \approx$ 6 nm of the tube can be evaluated by choosing the volume $\approx d^3$ comprising $10^3 - 10^4$ of $Xe$ atoms for 3-D thermodynamics to be meaningful. If the nanofluid is flat, it is described by the Onsager's theory,[27] otherwise, if $\ell \gg d$, the nanofluid is pencil shaped, the geometry which we are interested in. Maximal geometric length $\ell$ available for the fluid bounds the correlation length,[29]

$$\max_{T, \bar{n}} \{\xi\} = \ell. \qquad (3.21)$$

The situation $\xi = \ell$ is achieved exactly at the CP. At this state *all* the particles of the CNTs nanofluid are correlated providing the peculiarities of the equation of state.[27,29,30]

In the vicinity of the CP, the inverse susceptibility $|a_\ell|$ for the confined fluid is related with the bulk inverse susceptibility $|a|$ of Eq. (3.18) as[29]

$$|a_\ell|/|a| = \phi(\ell/\xi) \qquad (3.22)$$

where $\xi = \xi_0/\sqrt{|\tau|}$ Is the MF bulk correlation length. Eq. (3.22) can be further rewritten equivalently by using Eq. (3.18) and eliminating $\xi$ in favor of $\ell$,

$$|a_\ell| = \frac{k_B T_c}{x^2 \bar{n}_t} \left(\frac{\xi_0}{\ell}\right)^2 K^2(\delta), \qquad (3.23a)$$

where the function $K^2(\delta) = \delta \cdot \phi(\sqrt{\delta})$, $\delta = |\tau|(\ell/\xi_0)^2$, has the quadratic form near $\delta = 0$,

$$K(\delta) = 1 + \tfrac{1}{2}\delta^2 K''(0), \qquad (3.23b)$$

with $K''(0) > 0$, see Fig. 4. The $|a_\ell|$ of Eq. (3.23) is a special case of the $|a|$ of Eq. (3.18) when $\xi = \ell$ and $\tau = 0$. Confinement rigidity

$$c_\ell = \frac{k_B T_c}{x^2 \overline{n}_t} \xi_0^2 \qquad (3.24)$$

is chosen to be coinciding with the bulk rigidity $c$ of Eq. (3.20) irrespective of the length $\ell$ of the CNTs. We note, lastly, that the finite size correlation length is defined via the LG phenomenological parameters just as the bulk $\xi$ of Eq. (3.19), i.e.

$$\xi_\ell = \sqrt{c_\ell/|a_\ell|} = \ell/K(\delta) \qquad (3.25a)$$

$$\approx \ell\left(1 - \tfrac{1}{2}\delta^2 K''(0)\right), \quad \tau \to 0. \qquad (3.25b)$$

### D. Correlation function for CNTs nanofluid

We want to prove in this Section that the MF correlation function $\langle \delta n(\vec{r}_1) \delta n(\vec{r}_2) \rangle$ of the fluid within a closed container of the volume $\Omega$ is related with the Green function $G(\vec{r}_1, \vec{r}_2)$ of the operator $\hat{L} = (k_B T_c)^{-1}(|a_\ell| - c_\ell \Delta_{\vec{r}})$ as follows

$$\langle \delta n(\vec{r}_1) \delta n(\vec{r}_2) \rangle = G(\vec{r}_1, \vec{r}_2) - \frac{k_B T_c}{\Omega |a_\ell|}, \qquad (3.26a)$$

with the Green function satisfying the equation

$$\left(|a_\ell| - c_\ell \Delta_{\vec{r}_1}\right) G(\vec{r}_1, \vec{r}_2) = k_B T_c \delta(\vec{r}_1 - \vec{r}_2), \qquad (3.26b)$$

and conditioning by the zero flux at the boundary,

$$\nabla G(\vec{r}_1, \vec{r}_2)\big|_B = 0. \qquad (3.26c)$$

The boundary condition of Eq. (3.26c) is the necessary element for the sum rule of Eq. (C12) to be satisfied. In particular, the spatially homogeneous distribution of the perfect gas within the closed container obeys Eq. (3.26c). The averaging in Eq. (3.26a) is carried out over the Gaussian fluctuations of the density $(\delta n(\vec{r}) = \eta(\vec{r}))$ conditioned by $\int_\Omega d^3\vec{r}\, \delta n(\vec{r}) = 0$,

$$\langle \eta(\vec{r}_1) \eta(\vec{r}_2) \rangle = \int d\{\eta\}\, \eta(\vec{r}_1) \eta(\vec{r}_2) W(\{\eta(\vec{r})\}), \qquad (3.27a)$$

where the distribution

$$W(\{\eta(\vec{r})\}) = e^{-\delta F/k_B T} \delta\left(\int_\Omega d^3\vec{r}\, \eta(\vec{r})\right) \Big/ \int d\{\eta\} e^{-\delta F/k_B T} \delta\left(\int_\Omega d^3\vec{r}\, \eta(\vec{r})\right), \qquad (3.27b)$$

and the free energy fluctuation

$$\delta F/k_B T = \frac{1}{2k_B T}\int_\Omega d^3\vec{r}\,\eta(\vec{r})\bigl(|a_\ell|-c_\ell\Delta_{\vec{r}}\bigr)\eta(\vec{r}). \qquad (3.27c)$$

The $\delta$-function in Eq. (3.27b) singles out the functions $\eta(\vec{r})$ which satisfy the sum rule

$$\int_\Omega d^3\vec{r}\,\eta(\vec{r})=0, \qquad (3.28)$$

guarantying the conservation of the total number of atoms inside the confinement. If it were not the $\delta$-functions in Eq. (3.27b), the function $\langle \eta(\vec{r}_1)\eta(\vec{r}_2)\rangle$ would be exactly the Green function[27,39,67]

$$\langle \eta(\vec{r}_1)\eta(\vec{r}_2)\rangle = G(\vec{r}_1,\vec{r}_2). \qquad (3.29)$$

Proof of Eqs. (3.26), based on the Gaussian integrals,[67] we postpone until Appendix D where Eq. (3.29) is seen as a special case of the correlations in open systems with non-conserving Gaussian fluctuations of the density $\eta(\vec{r})$, see Eq. (D6). Eq. (3.26a) is the generalization of the Van Kampen's derivation[39] of OZ correlation function $\langle \delta n(\vec{r}_1)\delta n(\vec{r}_2)\rangle$ to the case of the closed containers of an arbitrary shape, especially for CNTs.

Two remarks are in order concerning Eq. (3.26a). First, the two terms in the r.-hand side of Eq. (3.26a) guarantee the sum rule of Eq. (C12) as it is obviously seen from the sum rule for the Green function of Eq. (D15). Second, the correlation function of Eq. (3.26a) has the long-range behavior due to the $\vec{r}$-independent constant $k_B T_c/(\Omega|a_\ell|)$. This long-range behavior comes from the global constraint of Eq. (3.28). Consider a function $\eta^0(\vec{r})$ obeying the sum rule of Eq. (3.28). Now deviate the magnitude of the function $\eta^0(\vec{r})$ only in two points $\vec{r}_1$ and $\vec{r}_2$ arbitrary distant from one another. If the new function at $\vec{r}_1$ is $\eta^0(\vec{r}_1)+\varepsilon$ then the new function at $\vec{r}_2$ should be $\eta^0(\vec{r}_2)-\varepsilon$ for the integral of Eq. (3.28) to keep zero. This implies the density fluctuations $\eta(\vec{r})$ inside the closed container to be correlated at an arbitrary separation $r=|\vec{r}_1-\vec{r}_2|$. If two points $\vec{r}_1$ and $\vec{r}_2$ are close to each other, $r\ll \xi, d$, the Green function[27,38] $G(\vec{r}_1,\vec{r}_2)=(k_B T_c/(4\pi c_\ell))e^{-r/\xi}/r$ governs the strength of the correlations, otherwise, if the points are far from the boundary and distant, $r\gg \xi, d$, they correlate with the strength

$$\langle \delta n(\vec{r}_1)\delta n(\vec{r}_2)\rangle \approx -k_B T_c/(\Omega|a_\ell|)=\mathrm{const}(\vec{r}_1,\vec{r}_2).$$

Before closing this Section we bring out the differences between the terms $k_B T_c/(\Omega|a_\ell|)$ belonging to the different geometries of the confinements such as the cylinders, discs, and cubes. By makng use $|a_\ell|$ of Eq. (3.23a), we show that the maximal term $k_B T_c/(\Omega|a_\ell|)$ is achieved for the cylindrical geometry of the container filled with the critical fluid. For 3-D long cylinders of the volume $\Omega_{\mathrm{tube}}=A\ell$ and the cross section area $A=\pi d^2/4$ we have

$$k_BT_c/(\Omega_{tube}|a_\ell|) = x^2\bar{n}_t \frac{1}{\xi_0^2 AK^2}\ell \sim \ell. \tag{3.30a}$$

For 3-D flat discs with the diameter $\ell$ (we specify by the symbol $\ell$ the largest geometrical length) and the thickness $h \ll \ell$, the inverse susceptibility at the CP is $|a_\ell| = (x^2\bar{n}_t)^{-1} k_BT_c (\xi_0/\ell)^2 K^2$. Here we use Eq. (3.23a) and set the correlation length equal to the maximal geometric length, $\xi = \ell$, in accord with Eq. (3.21). Using the volume $\Omega_{disc} = (\pi\ell^2/4)h$ we have

$$k_BT_c/(\Omega_{disc}|a_\ell|) = x^2\bar{n}_t \frac{1}{\xi_0^2 (\pi h/4) K^2} \sim \ell^0. \tag{3.30b}$$

For cubes of the volume $\Omega_{cube} = \ell^3$, Eq. (3.23a) reads $|a_\ell| = (x^2\bar{n}_t)^{-1} k_BT_c (\xi_0/\ell)^2 K^2$, hence

$$k_BT_c/(\Omega_{cube}|a_\ell|) = x^2\bar{n}_t \frac{1}{\xi_0^2 K^2}\frac{1}{\ell} \sim \ell^{-1}. \tag{3.30c}$$

If we set in Eqs. (3.30) the values $\xi_0 = h = d = 1$ nm and $K = 1$, then the maximal scaling $\sim \ell$ of the value $k_BT_c/(\Omega|a_\ell|)$ is achieved for the cylindrical critical fluid. Thus, it is appropriate to choose the cylindrical geometry of the critical fluid in order to get the large correlation function $\langle \delta n(\vec{r}_1)\delta n(\vec{r}_2)\rangle \sim \ell$ for the confined critical fluid. This correlation function drastically differs from conventional MF correlation function $\langle \delta n(\vec{r}_1)\delta n(\vec{r}_2)\rangle \sim \ell^{-1}$ for the critical fluids in the 3-D cubes.

Having derived the scaling relation for the coefficient $k_BT_c/(\Omega|a_\ell|)$ we have to derive the scaling relation for the Green function $G(\vec{r}_1,\vec{r}_2)$ entering into Eq. (3.26a). A slender 3-D cylindrical fluid is an archetypical 1-D system whose Green function can be calculated by averaging the exact Green function $G(\vec{r}_1,\vec{r}_2)$ of Eq. (3.26b) over the transverse coordinates of the cylinder, namely[68]

$$\langle G(\vec{r}_1,\vec{r}_2)\rangle_\perp = k_BT_c/(Ac_\ell) \cdot \frac{\xi_\ell}{2}\exp\left(-\frac{|z_1-z_2|}{\xi_\ell}\right) \sim \ell \tag{3.31}$$

near the CP where $\xi_\ell = \sqrt{c_\ell/|a_\ell|} \approx \ell$, see Eq. (3.25), and the factor $(k_BT_c/c_\ell)$ is the finite $\ell$-independent quantity by Eq. (3.20); see also Eqs. (E12), (E6). Green function of Eq. (3.31) meets the boundary condition of Eq. (3.26c) only with the exponential accuracy when one of the points $z_1$ or $z_2$ lies in the deep interior of the cylindrical fluid. In the next Section we derive the Green function $\langle G(\vec{r}_1,\vec{r}_2)\rangle_\perp$ conditioned exactly by Eq. (3.26c) for all the points $z_1$ and $z_2$ of the cylindrical fluid. However, even without this refinement, Eq. (3.31) enlightens the inherent scaling relationship between the Green function of the critical cylindrical fluid and the cylinder's length $\ell$. The scaling properties determined by Eqs. (3.31), (3.30a) are just the ones which guarantee the strong long range correlation function $\langle \delta n(\vec{r}_1) \delta n(\vec{r}_2) \rangle \sim \ell$ of Eq. (3.26a) or Eq. (3.3).

The identical scaling behavior for $\langle G(\vec{r}_1, \vec{r}_2) \rangle_\perp \sim \ell$ in Eq. (3.31) and for $k_B T_c / (\Omega_{tube} |a_\ell|) \sim \ell$ in Eq. (3.30a) provide the sum rule of Eq. (C12) for the function $\langle \delta n(\vec{r}_1) \delta n(\vec{r}_2) \rangle$ of Eq. (3.26a) as the $\ell$ goes to infinity. This implies the scaling $\langle G(\vec{r}_1, \vec{r}_2) \rangle_\perp \sim \ell$ to be consistent with the MF scaling $|a_\ell| \sim \ell^{-2}$ of Eq. (3.23a) justifying the MF ratio $\gamma/\nu = 2$ of the critical exponents for the critical fluids in the long CNTs.

### E. Projected Green function

Even without an algebraic bookkiping, the sum rule of Eq. (D15) places an important requirement on the Green function $G(\vec{r}_1, \vec{r}_2)$ of Eq. (3.26b),

$$G(\vec{r}_1, \vec{r}_2) = \frac{k_B T_c}{\Omega |a_\ell|} H(\vec{r}_1, \vec{r}_2), \quad (3.32)$$

where $H(\vec{r}_1, \vec{r}_2) \approx O(1)$ is a slow varying dimensionless function conditioned by $\int_\Omega H(\vec{r}_1, \vec{r}_2) d^3 \vec{r}_1 = \Omega$ so that the Green function scales as $G(\vec{r}_1, \vec{r}_2) \sim \ell$ due to the scaling factor $k_B T_c / (\Omega |a_\ell|) \sim \ell$ of Eq. (3.30a). The meaning of Eq. (3.32) is that all the particles of the correlation blob are equally correlated when $\xi_\ell \approx \ell$ so that the Green function $G(\vec{r}_1, \vec{r}_2)$ is almost spatially homogeneous one for most of the points $\vec{r}_1, \vec{r}_2$ in the long CNTs, see Fig. 1. In order to relate Eq. (3.32) with Eq. (3.31) we have used the relationship $|a_\ell| = c_\ell K^2 / \ell^2$ by Eqs. (3.23a), (3.24).

We prove in this Section an averaging theorem. The exact Green function $G(\vec{r}_1, \vec{r}_2)$ of Eq. (3.26b) as $\ell \gg d$ and $\ell \approx \xi_\ell$ asymptotically approaches to the $G$ averaged over the transversal coordinates of the cylinder,

$$G(\vec{r}_1, \vec{r}_2) \xrightarrow[\ell \approx \xi_\ell]{\ell/d \to \infty} \langle G(\vec{r}_1, \vec{r}_2) \rangle_\perp = \frac{k_B T_c}{\Omega |a_\ell|} K^2 \bar{H}(z_1, z_2) \quad (3.33a)$$

$$= \bar{G}(z_1, z_2), \quad (3.33b)$$

with dimensionless function

$$\bar{H}(z_1, z_2) = (2K \sinh(K))^{-1} \left( \cosh\left( K\left\{ 1 - \frac{z_1 + z_2}{\ell} \right\} \right) + \cosh\left( K\left\{ 1 - \frac{|z_1 - z_2|}{\ell} \right\} \right) \right), \quad (3.33c)$$

where the longitudinal coordinates $0 \leq z_1, z_2 \leq \ell$, the factor $k_B T_c/(\Omega |a_\ell|)$ is of Eq. (3.30a) and $K$ is given by Eq. (3.23b). Function $\bar{H}$ of Eqs. (3.33a, c) solves the equation

$$\left(\frac{K^2}{\ell^2} - \frac{\partial^2}{\partial z_1^2}\right)\bar{H}(z_1, z_2) = \frac{1}{\ell}\delta(z_1 - z_2),\qquad(3.34a)$$

subjected by the zero flux at the boundary,

$$\left.\frac{\partial \bar{H}}{\partial z_1}\right|_{z_1=0,\ell} = \left.\frac{\partial \bar{H}}{\partial z_2}\right|_{z_2=0,\ell} = 0.\qquad(3.34b)$$

From Eqs. (3.34) it follows the sum rule

$$\int_0^\ell dz_1 \bar{H} = \ell/K^2,\qquad(3.35)$$

hence, the integration of the both sides of Eq. (3.33a) over the volume $\Omega = A\ell$ of the cylinder gives the sum rule of Eq. (D15),

$$\int_A d\sigma \int_0^\ell dz_1 G(\vec{r}_1, \vec{r}_2) = k_B T_c/|a_\ell|.\qquad(3.36)$$

Notation $\bar{G}(z_1, z_2)$ in Eq. (3.33b) makes a convenient way of thinking about the function $\langle G(\vec{r}_1, \vec{r}_2)\rangle_\perp$ as depending on the $z$ coordinates only and we use the symbol $\langle \cdots \rangle_\perp$ which stands for the averaging of Eq. (3.26b) over the cross section at each longitudinal $z$ coordinate,

$$\langle G(\vec{r}_1, \vec{r}_2)\rangle_\perp = \frac{1}{A}\int_A dx_1 dy_1 G(\vec{r}_1, \vec{r}_2) = \bar{G}(z_1, z_2).\qquad(3.37)$$

Proof of Eqs. (3.33), (3.34) we postpone until Appendix E, and for the moment we are concerned with qualitative arguments in favor of Eqs. (3.33), (3.34). Rescale the coordinates of the the cylinder as $\{x' = x/d, y' = y/d, z' = z/\ell\}$ so that all the prime-coordinates lie between $0$ and $1$. Then multiplying both sides of Eq. (3.26b) on the factor $\ell^2/c_\ell$ leaves us with the equation

$$\left(K^2 - (\ell/d)^2 \left(\partial_{x_1'}^2 + \partial_{y_1'}^2\right) - \partial_{z_1'}^2\right)G = (k_B T_c/c_\ell)\ell d^{-2}\delta(\vec{r}_1' - \vec{r}_2'),\qquad(3.38)$$

where we have used the relationship $|a_\ell|/c_\ell = K^2/\ell^2$ between the paprameters $|a_\ell|$ and $c_\ell$ of Eq. (3.23a) and Eq. (3.24), respectively. The l.-hand side of Eq. (3.38) involves the large dimensionless parameter $(\ell/d)^2 \to \infty$ so that the Green function $G$ should be independend on the transverse coordinates $x_{1,2}'$, $y_{1,2}'$ ($x_i' \neq y_i'$), otherwise, the gradients in the $x'$ or $y'$ directions would not compensated by the comparable gradients in the $z'$ direction to get zero in the r.-hand side of Eq. (3.38) when $\vec{r}_1' \neq \vec{r}_2'$. We are forced to admit the asymptotic $\ell/d \to \infty$ of the Green function $G$ to be determined by the spatial homogenization of the Green function $G$ over the transverse coordinates $x'$ and $y'$. On applying Eq.(3.37) to Eqs. (3.26b, c) and invoking the Gauss's formula $\int_A dx_1 dy_1 \Delta_\perp G(\vec{r}_1, \vec{r}_2) = \int_B d\vec{l}\vec{\nabla} G = 0$ yield

$$\left((K/\ell)^2 - \partial_{z_1}^2\right)\bar{G}(z_1, z_2) = \frac{k_B T_c}{A c_\ell}\delta(z_1 - z_2). \qquad (3.39)$$

Eq. (3.39) yields straightforwardly Eq. (3.34a) upon replacement the function $\bar{G}$ with the function $\bar{H}$ by Eqs. (3.33a, b). We denote that another meaning of Eq. (3.37) is the coarse-grained Green function over the spatial scale of the size $d$ which is the minimal geometric size of the cylindes.

Having derived the asymptotic for the Green function of Eqs. (3.33) for the critical cylindrical fluid as $\ell/d \to \infty$, we are now in position to derive the asymptotic for the correlation function of Eq. (3.26a). Denote by the double brackets $\langle\langle \delta n(\vec{r}_1) \delta n(\vec{r}_2) \rangle\rangle = \langle\langle \delta n(\vec{r}_1) \delta n(\vec{r}_2) \rangle_{\delta n} \rangle_\perp$ the averaging of the function $\delta n(\vec{r}_1) \delta n(\vec{r}_2)$ over the Gaussian distribution $W(\{\delta n(\vec{r})\})$ of Eqs. (3.27b, c) followed by the spatial averaging over the transverse coordinates of the nanofluid. The later averaging is equivalent to the finding the function $\langle \delta n(\vec{r}_1) \delta n(\vec{r}_2) \rangle_{\delta n}$ in the asymptotic $\ell/d \to \infty$. By Eqs. (3.26a), (3.30a), and Eq. (3.33) we have

$$\langle\langle \delta n(\vec{r}_1) \delta n(\vec{r}_2) \rangle\rangle = \frac{k_B T_c}{\Omega |a_\ell|}\left(K^2 \bar{H}(z_1, z_2) - 1\right). \qquad (3.40)$$

By Eq. (3.40), the sum rule of Eq. (3.35) guarantees the sum rule of Eq. (C12). We stress here that for the density correlation function $\langle\langle \delta n(\vec{r}_1) \delta n(\vec{r}_2) \rangle\rangle$ be adequately dealt with, the two terms in the r.-hand side of Eq. (3.40) have to be taken into account. The correlation function of Eq. (3.40) for the closed containers should be compared with that for the open containers where the second summand of Eq. (3.40), i.e. $-k_B T_c / (\Omega |a_\ell|)$, ought to be dropped out leaving only the Green function in accord with Eq. (3.29).

We close this Section by quantifying how the cylindrical geometry of the confinement enhances the spatial coherence of the density fluctuations of the critical confined fluid. By Eqs. (3.40), (3.30a) at the CP for $K = 1$ and $x = 1$, we have $\langle\langle (\delta n)^2 \rangle\rangle \approx \bar{n}_t \ell / (A \xi_0^2) = N / \Omega_{\xi_0}^2$ or

$$\langle\langle (\delta N)_{\Omega_{\xi_0}}^2 \rangle\rangle \approx N, \qquad (3.41)$$

implying that, the fluctuation of the total number of atoms in the small volume $\Omega_{\xi_0} = A\xi_0$ is about the square root of the total number $N = \bar{n}_t A \ell$ of the atoms in the entire cylinder of the volume $\Omega = A\ell$.

### F. Levanyuk-Ginzburg criteria for CNTs nanofluid

Levananyuk-Ginzburg criteria[27]

$$\frac{1}{\Omega_\xi^2} \iint_{\Omega_\xi} \langle \delta n(\vec{r}_1) \delta n(\vec{r}_2) \rangle d^3\vec{r}_1 d^3\vec{r}_2 \ll \eta_*^2 = |a_\ell|/b_\ell \tag{3.42}$$

of the validity of the MF approximation at the CP can explicitly be calculated for the correlation function of Eq. (3.40). Exactly at the CP, we have $\xi_\ell = \ell$, see Eq. (3.21), so that the correlation volume $\Omega_\xi = A\xi_\ell$ of the cylindrical fluid exactly coincides with the volume of the tube, $\Omega_\xi = \Omega = A\ell$, and the concervation of the total number of atoms in the closed tube requires $\int_\Omega d^3\vec{r} \delta n(\vec{r}) = 0$. Hence, exactly at the CP the l.-hand side of Eq. (3.42) is zero, whereas the r.-hand side is the finite value $\eta_*^2 = (x\bar{n}_t)^2 (\xi_0/\ell)^2$. This value is the finite-size analog of the bulk value $\eta_*^2 = (x\bar{n}_t)^2 |\tau|$ in the MF approximation. It is easily to infer the prescription of how the $6$-fold integral in the left hand side of Eq. (3.42) approaches to zero when $\xi_\ell \to \ell$ if we carry out the calculations for a long toroidal tube. Due to the periodic conditions of the Green function of Eq. (E11) in the $z$-direction and the homogeneity of the Green function in the transversal directions of the toroidal tube, the l.-hand side of Eq. (3.42) is integrated to

$$\text{l.-h. side of Eq. (3.42)} = \frac{x^2 \bar{n}_t}{A\xi_0^2} \cdot \frac{\ell^2}{K^2 \xi_\ell} \left( \frac{1}{2} + \frac{\sinh\left(K\{\frac{\xi_\ell}{\ell} - \frac{1}{2}\}\right)}{2\sinh(K/2)} - \frac{\xi_\ell}{\ell} \right), \tag{3.43}$$

where we have used Eqs. (E9), (E11), and Eqs. (3.23), (3.40). Using the values of the parameters $K$ and $\xi_\ell$ from Eq. (3.23b) and Eq. (3.25b), respectively, we have

$$\text{l.-h. side of Eq. (3.42)} = C \frac{x^2 \bar{n}_t}{A\ell} (\ell/\xi_0)^2 \delta^2, \tag{3.44}$$

with constant $C = \left(1 - \frac{1}{2}\coth\left(\frac{1}{2}\right)\right) K''(0)/2 \approx O(1)$. The Ginzburg criterion is the ratio of the l.-hand side to the r.-hand side of Eq. (3.42), i.e.

$$\varepsilon_{Gi} = \bar{N}_t^{-1} \tau^2 (\ell/\xi_0)^7 C, \tag{3.45}$$

where $\delta$ is due to Eq. (3.23a) and $\bar{N}_t = A\ell\bar{n}_t$ is the total number of the atoms of the critical fluid in the toroidal tube. The $\varepsilon_{Gi} \ll 1$ is achieved when

$$\tau \ll \sqrt{\bar{N}_t} (\xi_0/\ell)^{7/2}, \tag{3.46}$$

i.e. the region of validity of the MF approximation encompasses the single critical point. We denote that the Ginzburg criterion places too strong relationships on the thermodynamic parameters at hand. The MF approximation is often applied even when the thermodynamic parameters lie outside of the validity of the Ginzburg criterion, i.e. when $\varepsilon_{Gi} \geq 1$.[36,37]

### G. Effective dipolar coupling $\bar{\upsilon}$

We use the asymptotic correlation function of Eq. (3.40) and the factor $k_B T_c / (\Omega |a_\ell|)$ of Eq. (3.30a) in order to calculate the effective coupling $\bar{\upsilon}$ of Eq. (3.2) as $\ell/d \to \infty$. By Eqs. (3.2) and (1.2) we have

$$\bar{\upsilon}_0 = \frac{1}{(A\ell)^2} \iint_\Omega d^3\vec{r}_1 d^3\vec{r}_2 \; \upsilon(\vec{r}_1, \vec{r}_2), \tag{3.47a}$$

and, in accord with Eq. (3.40), we decompose the fluctuation couplings $\bar{\upsilon}_\Delta$ into the two terms

$$\bar{\upsilon}_\Delta = \bar{\upsilon}_{\Delta_1} + \bar{\upsilon}_{\Delta_2}, \tag{3.47b}$$

$$\bar{\upsilon}_{\Delta_1} = \frac{1}{(A\ell)^2} \iint_\Omega d^3\vec{r}_1 d^3\vec{r}_2 \; \upsilon(\vec{r}_1, \vec{r}_2) \frac{\ell}{A\xi_0^2 \bar{n}_t} \tilde{H}(z_1, z_2), \tag{3.47c}$$

$$\bar{\upsilon}_{\Delta_2} = -\frac{1}{(A\ell)^2} \iint_\Omega d^3\vec{r}_1 d^3\vec{r}_2 \; \upsilon(\vec{r}_1, \vec{r}_2) \frac{\ell}{A\xi_0^2 \bar{n}_t K^2}, \tag{3.47d}$$

where we have used Eq. (3.30a) and the expression for the total number $N = x\bar{n}_t (A\ell)$ of spin carrying $^{129}Xe$ atoms with the concentration $\bar{n} = x\bar{n}_t$ of Eq. (3.15). It follows that the couplings $\bar{\upsilon}_{\Delta_1}$ and $\bar{\upsilon}_{\Delta_2}$ as well as the coupling $\bar{\upsilon}_0$ are independent of the fraction $x (\leq 1)$ of $^{129}Xe$ atoms in the entire solution with the critical number density $\bar{n}_t \approx 5 \text{ nm}^{-3}$ of $Xe$ atoms. Comparison of Eqs. (3.47c, d) with Eq. (3.47a) shows that the $\ell$ dependence of the couplings $\bar{\upsilon}_{\Delta_1}$, $\bar{\upsilon}_{\Delta_2}$ differ from that of the coupling $\bar{\upsilon}_0$ of Eq. (3.47a). In Eq. (3.47a), the factor before the double integral scales as $\ell^{-2}$ and after the spatial integral over $z_1$ we have[25] $\bar{\upsilon}_0 \sim \Omega^{-1} \sim \ell^{-1}$ since the integral over $0 \leq z_1 \leq \ell$ gives the additional factor $\ell$ and the integral over the relative coordinate $z_{12}$ is coupled with the integrand $\upsilon(\vec{r}_1, \vec{r}_2)$ yielding a constant. On this basis, the couplings $\bar{\upsilon}_{\Delta_1}$ and $\bar{\upsilon}_{\Delta_2}$ scale as a constant, i.e. $\bar{\upsilon}_{\Delta_1}$ and $\bar{\upsilon}_{\Delta_2} \sim \ell^0$, because of the additional factor $\ell$ in the integrands of Eqs. (3.47c, d). A special attention should be turned to the integrals in the expression $\bar{\upsilon}_{\Delta_1}$ of Eq. (3.47c). Qualitatively, one can regard the function $\tilde{H}(z_1, z_2)$ as almost a constant as in Eq. (3.32) so that the scaling $\bar{\upsilon}_{\Delta_1} \sim \ell^0$ holds. We postpone until Appendix F the derivation of the following expressions for the couplings of Eqs. (3.47),

$$\bar{\upsilon}_0 = 2\pi \frac{\gamma^2 \hbar^2}{A\ell}, \tag{3.48a}$$

$$\bar{\upsilon}_\Delta = \pi \frac{\gamma^2 \hbar^2}{(A\xi_0)^2 \bar{n}_t} \left( \frac{\coth(K)}{K} - \frac{1}{K^2} \right), \tag{3.48b}$$

for the long cylinders parallel to the external magnetic field $B_0$ along $z$ axis. The function $\Phi(K) = K^{-1}\coth(K) - K^{-2}$ monotone falls from $\Phi(0) = 1/3$ to zero as $\Phi(K) = 1/K$ for $K \gg 1$. Monotone dec-reasing function $\bar{\upsilon}_\Delta$ versus $K$ suggests that the the maximal $\bar{\upsilon}_\Delta$ occurs at the minimal $K$ which is realized at $T = T_c$ due to $|a_\ell|$ of Eq. (3.23a) and Fig. 4, see Fig. 6.

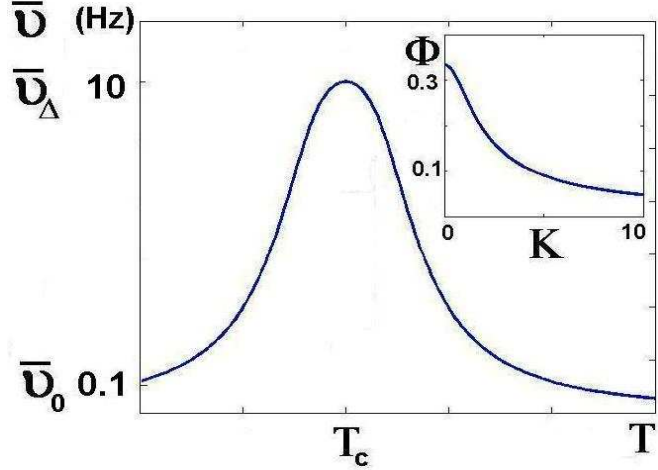

Fig. 6. Illustration of the function $\bar{\upsilon}_\Delta(T)$ when the parabolic curve $K(T)$ at $T = T_c$, see Eqs. (3.23a, b), is superimposed on the function $\bar{\upsilon}_\Delta(K)$ of Eqs. (3.48). The values $\bar{\upsilon}_0$ and $\bar{\upsilon}_\Delta$ refer to the example in the text below Eq. (3.49b). Insert shows the function $\Phi(K) = K^{-1}\coth(K) - K^{-2}$ of Eq. (3.48b).

Whereas the coupling $\bar{\upsilon}_0$ diminishes with the volume $A\ell$, the coupling $\bar{\upsilon}_\Delta$ is independent of the length $\ell$ of the tubes as we have elucidated in the scaling arguments after Eq. (3.47d). Due to $\ell$ elimination from the coupling $\bar{\upsilon}_\Delta$, the values $\ell$ and the cross section area $A$ enter into the expression $\bar{\upsilon}_\Delta$ of Eq. (3.48b) not on equial footing. The coupling $\bar{\upsilon}_\Delta$ is inversely proportional to the volume $A\xi_0$ squared, with $\xi_0$ being equal to the correlation length far from the CP, and inversely proportional to the the total con-centration $\bar{n}_t$ of $Xe$ atoms in the tubes. As Eqs. (3.48) stand, the effective dipolar interaction $\bar{\upsilon} = \bar{\upsilon}_0 + \bar{\upsilon}_\Delta$ does not averaged to zero if we enlarge the volume $\Omega$ of the tubes by enlarging theirs length $\ell$ keeping the cross section area $A = \pi d^2/4$ finite. As $\ell/d \to \infty$ the couplings $\bar{\upsilon}_0$ and $\bar{\upsilon}$ asymptotically become

$$\bar{\upsilon}_0 \xrightarrow{\ell/d \to \infty} 0, \tag{3.49a}$$

$$\bar{\upsilon} \xrightarrow{\ell/d \to \infty} \text{finite } \bar{\upsilon}_\Delta \neq 0. \tag{3.49b}$$

For example, put $A = 2$ nm$^2$, $\xi_0 = 0.2$ nm, $\ell = 10^3$ nm, $\bar{n}_t = 5$ nm$^{-3}$, and $K = 1$, then $(A\xi_0)^2 \bar{n}_t \approx 1$ nm$^3$ and $\bar{\upsilon}_\Delta/\bar{\upsilon}_0 \approx \frac{1}{3} \times 10^3$, i.e. all the effective dipolar interaction $\bar{\upsilon}$ of $^{129}Xe$ nuclei is due to the critical density fluctuations of $^{129}Xe$ atoms in the tubes. At these conditions, the coupling $\bar{\upsilon}_\Delta$ of Eq. (3.48b) is $\bar{\upsilon}_\Delta/(2\pi\hbar) \approx 10$ Hz in frequency units. This magnitude is just the one which we designated as $\bar{\upsilon}_\Delta \approx 10$ Hz and which we referred to as the typical effective coupling $\bar{\upsilon}_\Delta$ in Hz throughout Sections II.C-E.

The dependence of the coupling $\bar{v} = \bar{v}_0 + \bar{v}_\Delta$ on temperature $T$ is involved in the function $K(T)$ of Eq. (3.23b). Since the $K(T)$ run off to infinity as the temperature deviates from the critical temperature $T_c$, the coupling $\bar{v}_\Delta$ of Eq. (3.48b) tends to zero far from the CP, and we are left with

$$\bar{v} = \begin{cases} \bar{v}_\Delta = \pi\gamma^2\hbar^2\Phi(1)/\left((A\xi_0)^2 \bar{n}_t\right) \approx 10 \text{ Hz} & \text{- at the CP} \\ \bar{v}_0 = 2\pi\gamma^2\hbar^2/(A\ell) & \text{- far from the CP} \end{cases}. \quad (3.50)$$

The coupling $\bar{v} = \bar{v}_0 \sim 1/\Omega$ far from the CP can be assigned to the effective coupling $\bar{v}$ in the fluid with incoherent density correlations,[25] i.e. $\xi_\ell \ll \ell$ or $\langle \delta n(\vec{r}_1)\delta n(\vec{r}_2)\rangle = 0$, for $\vec{r}_1 \neq \vec{r}_2$. In the vicinity of the CP, the finite dipolar coupling $\bar{v}_\Delta \approx 10$ Hz is the only residual coupling due to its predominance over the term $\bar{v}_0 \sim 1/\Omega$ for the long cylindrical tubes. This predominance of the coupling $\bar{v}_\Delta$ over the $\bar{v}_0$ lies at the root of the enhancement of the solution NMR signals of the critical nanofluids in the long closed tubes described in Sections II.C-E.

## IV. CONCLUSION

We predicted the enhanced NMR intensities and the resonant frequency shift of $^{129}Xe$ critical nanofluid in the closed cylindrical nanotubes. This result may be interesting in view of the NMR experiments with supercritical $^{129}Xe$.[71] The enhanced signals rest on the observation that the spin carrying atoms of the critical nanofluid has the finite volume-independent residual dipolar coupling $\bar{v}_\Delta \approx 10$ Hz (specific to the $^{129}Xe$ fluid). The strong residual dipolar coupling $\bar{v}_\Delta \gg \bar{v}_0 \sim 1/\Omega$ is the result of the large-scale clustering which accelerates the flip-flops of the nuclear spins in the closed cylindrical critical nano-fluids. We discussed three NMR applications of the finite residual coupling $\bar{v}_\Delta$ of the critical hyperpolarized nanofluids: The FID with the large splitting $3\bar{v}_\Delta$ of the resonant frequencies, the power $\sim \bar{v}_\Delta^2 N^3$ of the spontaneous radiation that can exceed the Dicke's power, the absorption line shape $\chi''(\omega)$ with the large shift $\frac{3}{2}\bar{v}_\Delta N$ of the resonant frequency from the Larmor frequency, and appearing of the strong inherent dipolar field $\frac{3}{2}\bar{v}_\Delta N / |\gamma_{^{129}Xe}|$ that can exceed the external field $B_0$ of the strong magnets. One expects that the finite value of the coupling $\bar{v}_\Delta$ offer also the possibility of the fast rate $\sim \bar{v}_\Delta N$ of the polarization transfer between the nuclear spins of the critical hyperpolarized nanofluid and an electron spin which is in the thermodynamic equilibrium with the nanofluid and which is bound to the outer side-wall of the CNT.

The very generic relationships (1.1) – (1.3) for the effective interaction $\bar{v}$ and the large value of the correlation function $\langle \delta n(\vec{r}_1)\delta n(\vec{r}_2)\rangle \sim \ell$ of Eq. (3.3) for the critical cylindrical nanofluids are independent of the nature of interacting particles making up the nanofluids. We have discussed the enhanceement of the effective rate of the nuclear spin dynamics in the critical nanofluids, but one can explore, on the same footing, an inhaneement of the effective rates of the kinetically limited reactions in the critical confined solutions with the equilibrium large scale density fluctuations.[73] These drastic enhance-ments of the effective rates of the coherent kinetical processes contrast with the critical slowing down of the kinetical processes which one would expect in these circumstances.

**APPENDIX A: DERIVATION OF EQs. (2.40), (2.41)**

In deriving Eqs. (2.40), (2.41) from the FID $\langle I_x \rangle(t) = \text{tr}\{I_x \sigma(t)\}$ of Eq. (2.12) we, first, employ the density matrix $\sigma(t) = e^{-iH_0 t}\sigma(+0)e^{iH_0 t}$ where the state $\sigma(+0)$ is given by Eq. (2.39), the $H_0 = -\omega_0 I_z + \bar{H}_{dz}$ and the Hamiltonian $\bar{H}_{dz}$ is of Eq. (2.32), second, we use the commutator $[I_z, \bar{H}_{dz}] = 0$ and the identity $e^{-i\omega_0 I_z t} I_x e^{i\omega_0 I_z t} = I_x \cos(\omega_0 t) + I_y \sin(\omega_0 t)$, third, simplify the trace in the basis of $2^N$ orthonormal states that include the state $|0\rangle$, and finally, use the cyclic permutation of the operators within the trace, yielding

$$\langle I_x \rangle(t) = \cos(\omega_0 t)\langle 0|UI_x U^+|0\rangle + \sin(\omega_0 t)\langle 0|UI_y U^+|0\rangle, \tag{A1}$$

where we introduce the operator $U = e^{-i\frac{\pi}{2}I_Y} e^{i\bar{H}_{dz} t}$. To calculate the matrix element $\langle 0|UI_x U^+|0\rangle$ with the Hamiltonian $\bar{H}_{dz}$ of Eq. (2.32) we invoke the commutator $[\vec{I}^2, I_a] = 0$, $a = x, y, z$, and the identity[55]

$$e^{i\varphi I_z^2} I_+ e^{-i\varphi I_z^2} = e^{i\varphi(2I_z - 1)} I_+ \tag{A2}$$

with $\varphi = -\frac{3}{2}\bar{v}t$ in order to cast the matrix element $\langle 0|UI_x U^+|0\rangle$ into the form,

$$\langle 0|UI_x U^+|0\rangle = \tfrac{1}{2}\langle 0|e^{-i\frac{\pi}{2}I_Y}\left(e^{i\varphi(2I_z-1)}I_+ + h.c.\right)e^{i\frac{\pi}{2}I_Y}|0\rangle$$

$$= \text{Re}\left\{\langle 0|e^{i\varphi(2I_x - 1)}\left(-I_z + iI_y\right)|0\rangle\right\}, \tag{A3}$$

where we recalled the definition $I_x = (I_+ + I_-)/2$ and the following algebra of spin-$1/2$ operators, $e^{-i\frac{\pi}{2}I_Y} I_z e^{i\frac{\pi}{2}I_Y} = I_x$ and $e^{-i\frac{\pi}{2}I_Y} I_x e^{i\frac{\pi}{2}I_Y} = -I_z$. The Eq. (A3) can be put into the factorized form by taking advantage of the identity

$$\langle 0|e^{i\varphi(2I_x - 1)} = e^{-i\varphi}\prod_{n=1}^{N} \left({}_n\langle 0|\cos\varphi + i\, {}_n\langle 1|\sin\varphi\right), \tag{A4}$$

where we use the spin up states $|1\rangle_n = \begin{pmatrix}1\\0\end{pmatrix}_n$. Finally, we appeal to the properties

$$I_z|0\rangle = -\tfrac{N}{2}|0\rangle \quad \text{and} \quad iI_y|0\rangle = \tfrac{1}{2}\sum_{n=1}^{N} \underbrace{|...,1_n,...\rangle}_{\text{all other spins are down, }|0\rangle}. \tag{A5}$$

Eqs. (A4), (A5), once inserted in Eq. (A3), yield the real-valued matrix element

$$\langle 0|e^{i\varphi(2I_x-1)}(-I_z+iI_y)|0\rangle = \tfrac{N}{2}(\cos\varphi)^{N-1}. \tag{A6}$$

The matrix element $\langle 0|UI_yU^+|0\rangle$ of Eq. (A1) can be cast into the form of the imaginary part of the matrix element of Eq. (A6), namely

$$\langle 0|UI_yU^+|0\rangle = \text{Im}\{\langle 0|e^{i\varphi(2I_x-1)}(-I_z+iI_y)|0\rangle\} = 0. \tag{A7}$$

Eqs. (A1), (A6), and Eq. (A7) provide us with Eq. (2.41) of the main text.

### APPENDIX B: DERIVATION OF EQs. (2.61)

In the matrix element $\langle 0|I_xI_x(t')|0\rangle$ of Eq. (2.65), the operator $I_x(t') = e^{i\omega_0 I_z t'}\left(e^{-i\bar{H}_{dz}t'}I_x e^{i\bar{H}_{dz}t'}\right)e^{-i\omega_0 I_z t'}$ is simplified using the identity of Eq. (A2) and the identity $e^{i\omega_0 I_z t'}I_-e^{-i\omega_0 I_z t'} = I_-e^{-i\omega_0 t'}$, yielding

$$I_x(t') = \tfrac{1}{2}\left(I_-e^{i\varphi(2I_z-1)-i\omega_0 t'} + h.c.\right) \tag{B1}$$

with $\varphi = -\tfrac{3}{2}\bar{\upsilon}t'$. By Eq. (B1) and the identities $I_-|0\rangle = 0$, $\langle 0|I_+ = 0$, the matrix element of Eq. (2.65) becomes

$$\langle 0|I_xI_x(t')|0\rangle = \tfrac{1}{4}\langle 0|I_-e^{-2i\varphi I_z}I_+|0\rangle e^{i\varphi+i\omega_0 t'}. \tag{B2}$$

For the state $|0\rangle = |0_1,\ldots,0_N\rangle$, $I_z|0\rangle = -\tfrac{N}{2}|0\rangle$, $e^{2i\varphi I_z}|0\rangle = e^{-i\varphi N}|0\rangle$, and exploiting the property

$$I_+|0\rangle = \sum_{n=1}^{N}\underbrace{|\ldots,1_n,\ldots\rangle}_{\text{all other spins are down, }|0\rangle} \tag{B3}$$

the matrix element of Eq. (B2) takes the form

$$\langle 0|I_xI_x(t')|0\rangle = \tfrac{N}{4}e^{-i\varphi+i\omega_0 t'+i\varphi N} = \tfrac{N}{4}e^{i\omega_* t'} \tag{B4}$$

with $\omega_* = \omega_0 - \tfrac{3}{2}\bar{\upsilon}(N-1)$. After inserting Eq. (B4) in Eq. (2.65) and performing the integral $\int_0^\infty dt'\sin(\omega t')\sin(\omega_* t') = \tfrac{\pi}{2}(\delta(\omega-\omega_*) - \delta(\omega+\omega_*))$, see Eq. (2.45), we arrive at Eqs. (2.61).

### APPENDIX C: THE DENSITY CORRELATION FUNCTIONS. THE SUM RULES

The microscopic density $n(\vec{R}) = \sum_{\nu=1}^{N}\delta(\vec{R}-\vec{r}_\nu)$ of Eq. (1.4) is the stochastic function to the extent of the irregularity of the coordinates $\vec{r}_\nu$ of the $N$ atoms within the confinement. The ensemble average of the function $n(\vec{R})$ is related with the 1-point reduced distribution function (RDF)

$$\rho_{c\ell}^{(1)}(\vec{r}_1) = \int d\vec{r}_2...d\vec{r}_{N+M}\rho_{c\ell}(\vec{r}^{(N+M)}) \tag{C1}$$

as follows[64]

$$\langle n(\vec{R})\rangle = \int d\vec{r}^{(N+M)}\rho_{c\ell}(\vec{r}^{(N+M)})n(\vec{R})$$

$$= \sum_{\nu=1}^{N}\int d\vec{r}_\nu \int d(\vec{r}^{(N+M)}/\vec{r}_\nu)\rho_{c\ell}(\vec{r}^{(N+M)})\delta(\vec{R}-\vec{r}_\nu) = N\rho_{c\ell}^{(1)}(\vec{R}), \tag{C2}$$

where the symbol $d(\vec{r}^{(N+M)}/\vec{r}_\nu)$ denotes the differential over all the $(N+M)$ coordinates except the coordinate $\vec{r}_\nu$, in addition, the $(N+M)$ particle distribution function $\rho_{c\ell}(\vec{r}^{(N+M)})$ is assumed to be fully symmetric with respect to any interchanging of the coordinates $\vec{r}_1,...,\vec{r}_N$ of the $N$ identical spin carrying atoms $^{129}Xe$ as well as fully symmetric over the remaining $M$ coordinates of atoms $^{132}Xe$. If $Xe$ atoms were independent, the 1-particle RDF would be $\rho_{c\ell}^{(1)}(\vec{R}) = 1/\Omega$. For a homogeneous system of interact-ting atoms, the equality

$$\langle n(\vec{R})\rangle = N/\Omega \tag{C3}$$

holds for any point $\vec{R}$. The ensemble averaging

$$\langle n(\vec{R}_1)n(\vec{R}_2)\rangle = \int d\vec{r}^{(N+M)}\rho_{c\ell}(\vec{r}^{(N+M)})n(\vec{R}_1)n(\vec{R}_2) \tag{C4}$$

of the 2-point microscopic density

$$n(\vec{R}_1)n(\vec{R}_2) = \sum_{\substack{\nu_1=1,\nu_2=1 \\ \nu_1 \neq \nu_2}}^{N}\delta(\vec{R}_1-\vec{r}_{\nu_1})\delta(\vec{R}_2-\vec{r}_{\nu_2}) + \sum_{\nu=1}^{N}\delta(\vec{R}_1-\vec{r}_\nu)\delta(\vec{R}_1-\vec{r}_\nu) \tag{C5}$$

is derived[64] in a similar fashion in terms of the 1-point RDF of Eq. (C1) and the 2-point RDF $\rho_{c\ell}^{(2)}(\vec{r}_1,\vec{r}_2)$ of Eq. (2.34),

$$\langle n(\vec{R}_1)n(\vec{R}_2)\rangle = N(N-1)\rho_{c\ell}^{(2)}(\vec{R}_1,\vec{R}_2) + N\delta(\vec{R}_1-\vec{R}_2)\rho_{c\ell}^{(1)}(\vec{R}_1). \tag{C6}$$

For $\vec{R}_1 \neq \vec{R}_2$, we arrive at the relationship[64,27]

$$\langle n(\vec{R}_1)n(\vec{R}_2)\rangle = N(N-1)\rho_{c\ell}^{(2)}(\vec{R}_1,\vec{R}_2) \tag{C7}$$

that is used in obtaining Eq. (3.1) from Eq. (2.33). Except the average density $\langle n(\vec{R})\rangle = N/\Omega$ of Eq. (C3), the stochastic function $n(\vec{R})$ is endowed with the fluctuation around the average density,

$$\delta n(\vec{R}) = n(\vec{R}) - \langle n(\vec{R})\rangle, \quad \text{with} \quad \langle \delta n(\vec{R})\rangle = 0. \tag{C8}$$

Accordingly, for the homogeneous solution and $\vec{R}_1 \neq \vec{R}_2$, we have

$$\langle n(\vec{R}_1) n(\vec{R}_2) \rangle = (N/\Omega)^2 + \langle \delta n(\vec{R}_1) \delta n(\vec{R}_2) \rangle. \tag{C9}$$

By Eq. (C9) and Eq. (C7), we have for $N \gg 1$ and $\vec{R}_1 \neq \vec{R}_2$,

$$\rho_{c\ell}^{(2)}(\vec{R}_1, \vec{R}_2) = (1/\Omega)^2 + \langle \delta n(\vec{R}_1) \delta n(\vec{R}_2) \rangle / N^2. \tag{C10}$$

If the correlations were absent, $\langle \delta n(\vec{R}_1) \delta n(\vec{R}_2) \rangle = 0$, the $\rho_{c\ell}^{(2)}(\vec{R}_1, \vec{R}_2) = (1/\Omega)^2$. The sum rules follow from the definition $n(\vec{R})$ of Eq. (1.4) and the normalization to unity of the function $\rho_{c\ell}(\vec{r}^{(N+M)})$ of Eq. (2.8),

$$\int_\Omega d^3\vec{R} \langle n(\vec{R}) \rangle = N,$$

$$\int_\Omega d^3\vec{R}_1 \langle n(\vec{R}_1) n(\vec{R}_2) \rangle = N \langle n(\vec{R}_2) \rangle. \tag{C11}$$

In terms of the fluctuations $\delta n(\vec{R})$ of Eq. (C8), the sum rules of Eqs. (C11) read

$$\int_\Omega d^3\vec{R} \langle \delta n(\vec{R}) \rangle = 0,$$

$$\int_\Omega d^3\vec{R}_1 \langle \delta n(\vec{R}_1) \delta n(\vec{R}_2) \rangle = 0. \tag{C12}$$

**APPENDIX D: DERIVATION OF EQs.(3.26)**

The statistical averages in Eqs. (3.27) are derivable from the Gaussian integral

$$\frac{\int d\{\eta\} \exp\left(-E(\{\eta(\vec{r})\}) - i\int_\Omega d^3\vec{r}\, \omega(\vec{r})\eta(\vec{r})\right)}{\int d\{\eta\} \exp\left(-E(\{\eta(\vec{r})\})\right)} = \exp\left(-\frac{1}{2}\iint_\Omega d^3\vec{r}_1 d^3\vec{r}_2\, \omega(\vec{r}_1) G(\vec{r}_1,\vec{r}_2) \omega(\vec{r}_2)\right), \tag{D1}$$

where

$$L_{\vec{r}_1\vec{r}_2} = \langle \vec{r}_1 | \hat{L} | \vec{r}_2 \rangle = (k_B T_c)^{-1} \delta(\vec{r}_1 - \vec{r}_2)(|a_\ell| - c_\ell \Delta_{\vec{r}_1}) \tag{D2}$$

is $\vec{r}_1$-$\vec{r}_2$ element of the operator $\hat{L} = (k_B T_c)^{-1}(|a_\ell| - c_\ell \Delta_{\vec{r}_1})$ in the $|\vec{r}\rangle$-representation,[67] and we introduced the functional

$$E(\{\eta(\vec{r})\}) = \frac{1}{2}\iint_\Omega d^3\vec{r}_1 d^3\vec{r}_2\, \eta(\vec{r}_1) L_{\vec{r}_1\vec{r}_2} \eta(\vec{r}_2), \tag{D3}$$

and the function

$$G(\vec{r}_1, \vec{r}_2) = \langle \vec{r}_1 | \hat{L}^{-1} | \vec{r}_2 \rangle \tag{D4}$$

is equal to the $\vec{r}_1$ - $\vec{r}_2$ element of the operator $\hat{L}^{-1} = k_B T_c \left( |a_\ell| - c_\ell \Delta_{\vec{r}_1} \right)^{-1}$. The $G(\vec{r}_1, \vec{r}_2)$ of Eq. (D4) is the Green function of the operator $\hat{L}$ of Eq. (3.26b) as it is easily seen from the identity

$$\int_\Omega d^3\vec{r}_3 \langle \vec{r}_1 | \hat{L} | \vec{r}_3 \rangle \langle \vec{r}_3 | \hat{L}^{-1} | \vec{r}_2 \rangle = \langle \vec{r}_1 | \vec{r}_2 \rangle = \delta(\vec{r}_1 - \vec{r}_2). \tag{D5}$$

Eq. (D1) gives the formal expression for the Green function,

$$\langle \eta(\vec{r}_1)\eta(\vec{r}_2) \rangle = -\frac{1}{J(\{0\})} \frac{\delta^2 J(\{\omega(\vec{r})\})}{\delta\omega(\vec{r}_1)\delta\omega(\vec{r}_2)} \bigg|_{\omega(\vec{r})=0} = G(\vec{r}_1, \vec{r}_2), \tag{D6}$$

Where the $J$ - integral is

$$J(\{\omega(\vec{r})\}) = \int d\{\eta\} \exp\left( -E(\{\eta(\vec{r})\}) - i\int_\Omega d^3\vec{r}\, \omega(\vec{r})\eta(\vec{r}) \right). \tag{D7}$$

The function $\langle \eta(\vec{r}_1)\eta(\vec{r}_2) \rangle$ of Eqs. (3.26), (3.27) reads,

$$\langle \eta(\vec{r}_1)\eta(\vec{r}_2) \rangle = -\frac{1}{J_\delta(\{0\})} \frac{\delta^2 J_\delta(\{\omega(\vec{r})\})}{\delta\omega(\vec{r}_1)\delta\omega(\vec{r}_2)} \bigg|_{\omega(\vec{r})=0} \tag{D8}$$

where we introduced the $J_\delta$ - integral

$$J_\delta(\{\omega(\vec{r})\}) = \int D\{\eta\} \exp\left( -E(\{\eta(\vec{r})\}) - i\int_\Omega d^3\vec{r}\, \omega(\vec{r})\eta(\vec{r}) \right) \delta\left( \int_\Omega d^3\vec{r}\, \eta(\vec{r}) \right), \tag{D9}$$

with the differential element

$$D\{\eta\} = d\{\eta\} \bigg/ \int D\{\eta\} \exp\left( -E(\{\eta(\vec{r})\}) \right). \tag{D10}$$

The normalization factor of Eq. (D10) cancels out in calculating Eq. (D8), however, this normalization is very convenient to achieve agreement with the Gaussian integral of Eq. (D1) in what follows. We want to derive an explicit expression for the $J_\delta$ - integral of Eq. (D9) by representing the $\delta$ -function in Eq. (D9) via its Fourier transform $\delta\left( \int_\Omega d^3\vec{r}\, \eta(\vec{r}) \right) = \int_{-\infty}^{\infty} \frac{dQ}{2\pi} \exp\left( -iQ \int_\Omega d^3\vec{r}\, \eta(\vec{r}) \right)$, yielding

$$J_\delta(\{\omega(\vec{r})\}) = \int_{-\infty}^{\infty} \frac{dQ}{2\pi} \exp\left( -\frac{1}{2} \iint_\Omega d^3\vec{r}_1 d^3\vec{r}_2 \left( \omega(\vec{r}_1) + Q \right) G(\vec{r}_1, \vec{r}_2) \left( \omega(\vec{r}_2) + Q \right) \right), \tag{D11}$$

where we make use the Gaussian integral of Eq. (D1). It remains to calculate the Gaussian integral over the $Q$ -variable in Eq. (D11) resulting in

$$J_\delta(\{\omega(\vec{r})\}) = (2\pi)^{-1}\sqrt{\pi/\alpha}\exp(\sigma), \tag{D12}$$

where the exponent

$$\sigma(\{\omega(\vec{r})\}) = \frac{\beta^2}{4\alpha} - \gamma \tag{D13}$$

involves the coefficient $\alpha$ and the functionals $\beta$ and $\gamma$,

$$\alpha = \tfrac{1}{2}\iint_\Omega d^3\vec{r}_1 d^3\vec{r}_2\, G(\vec{r}_1, \vec{r}_2)_2 = \tfrac{1}{2}\Omega(k_B T_c/|a_\ell|),$$

$$\beta = \iint_\Omega d^3\vec{r}_1 d^3\vec{r}_2\, G(\vec{r}_1, \vec{r}_2)\omega(\vec{r}_1) = (k_B T_c/|a_\ell|)\int_\Omega d^3\vec{r}_1\omega(\vec{r}_1),$$

$$\gamma = \tfrac{1}{2}\iint_\Omega d^3\vec{r}_1 d^3\vec{r}_2\, \omega(\vec{r}_1)G(\vec{r}_1,\vec{r}_2)\omega(\vec{r}_2). \tag{D14}$$

In obtaining Eqs. (D14) we used, first, Eq. (3.26b) for the Green function $G(\vec{r}_1, \vec{r}_2)$, second, the Gauss's formula $\int_\Omega \Delta G d^3\vec{r} = \int_B \vec{\nabla}G d\vec{s} = 0$ relating the volume integral with the surface integral, third, the bounndary condition of Eq. (3.26c), fourth, the sum rule

$$\int_\Omega d^3\vec{r}_1 G(\vec{r}_1,\vec{r}_2) = k_B T_c/|a_\ell|, \tag{D15}$$

and, finally, the symmetry property $G(\vec{r}_1,\vec{r}_2) = G(\vec{r}_2,\vec{r}_1)$.

Since $\sigma(\{0\}) = 0$ and $\delta\sigma(\{\omega(\vec{r})\})/\delta\omega(\vec{r}_1)\big|_{\omega=0} = 0$, straightforward differentiation of Eq. (D8) by applying the rule $\delta\omega(\vec{r}_1)/\delta\omega(\vec{r}_2) = \delta(\vec{r}_1 - \vec{r}_2)$ to Eqs. (D12)-(D14) gives

$$\langle\eta(\vec{r}_1)\eta(\vec{r}_2)\rangle = -\frac{\delta^2\sigma(\{\omega(\vec{r})\})}{\delta\omega(\vec{r}_1)\delta\omega(\vec{r}_2)}\bigg|_{\omega(\vec{r})=0} = G(\vec{r}_1,\vec{r}_2) - \frac{k_B T_c}{\Omega|a_\ell|}. \tag{D16}$$

By Eq. (D16) the sum rule of Eq. (D15) guarantees the sum rule of Eq. (C12), with $\eta(\vec{r}) = \delta n(\vec{r})$.

**APPENDIX E: DERIVATION OF EQs. (3.33), (3.34)**

**ASYMPTOTIC GREEN FUNCTION**

It is simpler to get the asymptotic of Eqs. (3.33) as $\ell/d \to \infty$ for a box with the length $\ell$ and the rectangular cross section of the side $d$. Then we generalize the derivation to the Green function for an arbitrary cross section.

The Green function of Eqs. (3.26b, c) for the box reads[68]

$$G(\vec{r}_1,\vec{r}_2) = (k_B T_c/c_\ell) \sum_{k,l,m=0}^{\infty} \frac{\psi^*_{k,l,m}(\vec{r}_1)\psi_{k,l,m}(\vec{r}_2)}{\lambda_{k,l,m}}, \tag{E1}$$

where the complete set of the orthonormalized functions is

$$\psi_{k,l,m}(\vec{r}_1) = N_k N_l N_m \cos(x_1 \pi k/d)\cos(y_1 \pi l/d)\cos(z_1 \pi m/\ell), \tag{E2}$$

with $N_k = 1/\sqrt{d}$ for $k=0$ and $N_k = \sqrt{2/d}$ for $k=1,2,...,\infty$, analogously for $N_l$, and $N_m = 1/\sqrt{\ell}$ for $m=0$ and $N_m = \sqrt{2/\ell}$ for $m=1,2,...,\infty$. The eigenvalues are

$$\lambda_{k,l,m} = \left(\frac{K}{\ell}\right)^2 + \left(\frac{\pi k}{d}\right)^2 + \left(\frac{\pi l}{d}\right)^2 + \left(\frac{\pi m}{\ell}\right)^2, \tag{E3}$$

where we have used $|a_\ell|/c_\ell = K^2/\ell^2$ by Eqs. (3.23a), (3.24). The eigenfunctions of Eq. (E2) and eigenvalues of Eq. (E3) solve the equation

$$\left((K/\ell)^2 - \Delta\right)\psi_{k,l,m}(\vec{r}) = \lambda_{k,l,m}\psi_{k,l,m}(\vec{r}) \tag{E4}$$

with the boundary condition $\nabla\psi_{k,l,m}(\vec{r})\big|_B = 0$. Each summand of Eq. (E1) can be represented, up to the factor $(k_B T_c/c_\ell)$, as

$$\ell^2 \frac{\psi^*_{k,l,m}(\vec{r}_1)\psi_{k,l,m}(\vec{r}_2)}{K^2 + (\ell/d)^2 \pi^2(k^2+l^2)+\pi^2 m^2}, \tag{E5}$$

with the common factor $\ell^2$ for all the terms in the sum of Eq. (E1). When $\ell/d \to \infty$ all the terms of Eq. (E1) belonging to $k,l = 1,2,...,\infty$ vanish, leaving the dominant term with $k=l=0$. The quantum numbers $k=l=0$ are responsible for spatially homogeneous ('ground') state over the transverse coordinates $x, y$ of the box. Using the normalization $N^2_{k=0} = N^2_{l=0} = 1/d$, we have sought asymptotic of the Green function for the box,

$$G(\vec{r}_1,\vec{r}_2) = \frac{k_B T_c}{c_\ell d^2}\tilde{H}(z_1,z_2), \tag{E6}$$

where the function $\tilde{H}$ stands for the sum over the integer $m$,

$$\tilde{H}(z_1,z_2) = \sum_{m=0}^{\infty} N_m^2 \frac{\cos\left(\frac{\pi m}{\ell}z_1\right)\cos\left(\frac{\pi m}{\ell}z_2\right)}{(K/\ell)^2 + (\pi m/\ell)^2}. \tag{E7}$$

The sum of Eq. (E7) is simplified using the standard sum[70]

$$\sum_{m=1}^{\infty}\frac{\cos(n\theta)}{m^2+\varsigma^2}=-\frac{1}{2\varsigma^2}+\frac{\pi}{2\varsigma}\frac{\cosh(\varsigma\{\pi-|\theta|\})}{\sinh(\pi\varsigma)}, \quad (E8)$$

yielding

$$\tilde{H}(z_1,z_2)=\ell\bar{H}(z_1,z_2), \quad (E9)$$

where the function $\bar{H}(z_1,z_2)$ is exactly Eq. (3.33c). Combining together Eqs. (E9), (E6) as well as using the volume $\Omega=\ell d^2$ and the relationship $c_\ell=|a_\ell|\ell^2/K^2$, we arrive at Eq. (3.33). We denote that Eqs. (E7), (E9) guarantee the function $\bar{H}(z_1,z_2)$ to be the Green function obeying Eq. (3.34a).

For the toroidal tube,[60,71] the function

$$\tilde{H}(z_1,z_2)=\sum_{m=0}^{\infty}\frac{1}{\ell}\frac{\exp(i2\pi m(z_2-z_1)/\ell)}{(K/\ell)^2+(2\pi m/\ell)^2} \quad (E10)$$

is the analog of the function $\tilde{H}$ of Eq. (E7) for the case of the tube translationally invariant in $z$-direction. By Eq. (E8), the Eq. (E10) is given explicitly in the form

$$\tilde{H}(z_1,z_2)=\ell\frac{\cosh\left(K\{\tfrac{1}{2}-\tfrac{|z_2-z_1|}{\ell}\}\right)}{2K\sinh(K/2)}. \quad (E11)$$

This function depends on the coordinates as $|z_1-z_2|$ and is very convenient in calculating the spatial integrals in the Appendix F. In order to get 1-D Green function for the bulk, $\ell\to\infty$, one should simply represent Eq. (E11) in the form

$$\tilde{H}(z_1,z_2)=\frac{\ell}{2K}\left(e^{-K\frac{|z_2-z_1|}{\ell}}+\frac{2\cosh(K(z_2-z_1)/\ell)}{e^K-1}\right),$$

next replace the ratio $\ell/K$ with $\xi_\ell$, since $\ell$ is now the silent variable for infinitely large diameter $\ell/\pi$ of the torus with $\ell\gg\xi_\ell$, then we drop the second summand and come to old-fashioned 1-D Green function

$$\tilde{H}(z_1,z_2)=\frac{\xi_\ell}{2}\exp\left(-\frac{|z_1-z_2|}{\xi_\ell}\right), \quad (E12)$$

yielding Eq. (3.31).

Asymptotic of the Green function of Eqs. (3.26b, c) for cylinder is calculated a lit bit cumbersome than that for the box. Let $d$ be the diameter of the cylinder with length $\ell$. The particular functions in which we expand the Green function $G$ of Eq. (E1) are the orthonormalized eigenfunctions of Eq. (E4) for cylinder.[68,72] Designate by

$$\psi_{k=0,l=0,m}(\rho,\varphi,z) = \left(1/\sqrt{A}\right) N_{0,0,m} \cos(z\pi m/\ell) \tag{E13}$$

the orthonormalized eigenfunctions that are spatially homogeneous over the transverse coordinates $\rho$, $\varphi$ of the cylinder with $A = \pi d^2/4$ standing for the cross section area of the cylinder and the normalization $N_{0,0,m}$ being equal to the $N_m$ of Eq. (E2). We show that the functions of Eq. (E13) are central to the problem of finding the asymptotic Green function whereas the eigenfunctions which are inhomogeneous in the $\{\rho,\varphi\}$ plane, namely[72]

$$\psi_{k,l,m}(\rho,\varphi,z) = N_{k,l,m} e^{il\varphi} J_l\left(\mu_{k,l}\,\rho/(d/2)\right) \cos(z\pi m/\ell) \tag{E14}$$

can be eliminated in the limit $\ell/d \to \infty$, with indices running over $k = 1,2,\ldots,\infty$, $l = 0,1,\ldots,\infty$, and $m = 0,1,\ldots,\infty$. In Eq. (E14), the positive roots $\mu_{k,l} > 0$ $(\mu_{k,l} \neq 0)$ solve the equation

$$\left.\frac{d}{d\mu} J_l(\mu)\right|_{\mu=\mu_{k,l}} = 0 \tag{E15}$$

for the Bessel functions $J_l(\mu)$ of the first kind. The eigenvalues of Eq. (E4) are[72]

$$\lambda_{k,l,m} = \left(\frac{K}{\ell}\right)^2 + \left(\frac{\mu_{k,l}}{(d/2)}\right)^2 + \left(\frac{\pi m}{\ell}\right)^2. \tag{E16}$$

Eigenfunctions $\psi_{0,0,m}$ of Eq. (E13) can be associated with the eigenvalues $\lambda_{0,0,m}$ of Eq. (E16) with

$$\mu_{0,0} = 0. \tag{E17}$$

Having got all the eigenfunctions $\psi_{k,l,m}$ and eigenvalues $\lambda_{k,l,m}$ of Eq. (E13), each term of the Green function of Eq. (E1) can be represented, except for the factor $(k_B T_c/c_\ell)$, as

$$\ell^2 \frac{\psi^*_{k,l,m}(\vec{r}_1)\psi_{k,l,m}(\vec{r}_2)}{K^2 + \left(\ell/(d/2)\right)^2 \mu_{k,l}^2 + \pi^2 m^2} \tag{E18}$$

with the common factor $\ell^2$ for all the terms in the sum of Eq. (E1). As $\ell/d \to \infty$ the asymptotic of the ratio in Eq. (E18) is dominated by the terms with $\mu_{k=0,l=0} = 0$, i.e. by the terms of Eq. (E13) standing for spatially homogeneous functions $\psi_{0,0,m}$ over the $\rho,\varphi$ coordinates. All the other terms in Eq. (E18) belonging to $k = 1,2,\ldots,\infty$ and $l = 0,1,\ldots,\infty$ have $\mu_{k,l} > 0$, see Eq. (E15), so that all the corresponding summands in the Green function of Eq. (E1) vanish in the limit $\ell/d \to \infty$. Thus, the Green function of Eq. (E1) as $\ell/d \to \infty$ asymptotically becomes

$$G(\vec{r}_1,\vec{r}_2) \xrightarrow{\ell/d \to \infty} \overline{G}(z_1, z_2) = (k_B T_c/c_\ell) \sum_{m=0}^{\infty} \frac{\psi^*_{0,0,m}(\vec{r}_1)\psi_{0,0,m}(\vec{r}_2)}{(K/\ell)^2 + (\pi m/\ell)^2} \quad . \tag{E19}$$

Eq. (E19) coincides with the asymptotic $G(\vec{r}_1,\vec{r}_2)$ of Eqs. (E6), (E7) for the box in which we should simply replace $d^2$ with $A = \pi d^2/4$.

### APPENDIX F: DERIVATION OF EQs. (3.48)

The 6-th fold integrals in Eqs. (1.2), (1.3) are the non-zero quantities as compared with the zero integrals in the sum rules of Eqs. (C12). We break the calculation of the 6 -th fold integrals in Eqs. (3.47) into 6 steps.

Step 1: Introduce the independent coordinates $\vec{r}_1$ and $\vec{r}_{21} = \vec{r}_2 - \vec{r}_1$. The transformation $\{\vec{r}_1, \vec{r}_2\} \to \{\vec{r}_1, \vec{r}_{21}\}$ has the Jacobian of unity.

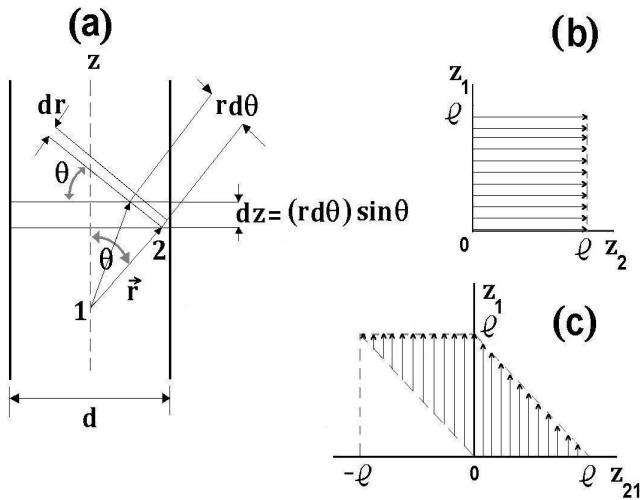

Fig. 7. Illustrations to Appendix F . (a) shows step 2, (b) and (c) correspond to steps (4) and (5). In figure (c), the up arrows point to the new direction of integration in Eq. (F1).

Step 2: Introduce the coordinates $\{r, \varphi, z\}$ where $r$ stands for the radius in the spherical coordinates, $\varphi$ is the azimuthal angle, and $z = r \cos \theta$ with $\theta$ being the polar angle, see Fig. 7 (a). The differential volume element $d^3\vec{r} = r^2 \sin \theta d\theta d\varphi dr$ of the spherical coordinates is transformed to $d^3\vec{r} = rdz d\varphi dr$ in the new coordinate system $\{r, \varphi, z\}$.

Step 3: We devide the cylinder into the stack of the disks of width $dz$ and the diameter $d$. Pick out a point $\vec{r}_1$ on the axis of the cylinder so that, at each fixed $z_{21}$, a point $\vec{r}_2$ lies on a disk with a radius $r_{21}$ which is in the limits $|z_{21}| \le r_{21} \le \sqrt{z_{21}^2 + (d/2)^2}$, $0 \le d\varphi_{21} \le 2\pi$. A point $\vec{r}_2$ with $r_{21} = |z_{21}|$ is on the axis of the cylinder with $\theta_{21} = 0$ and a point $\vec{r}_2$ with $r_{21} = \sqrt{z_{21}^2 + (d/2)^2}$ lies on the wall of the cylinder.

Step 4: Since $0 \leq z_1 \leq \ell$ and $0 \leq z_2 \leq \ell$, at each fixed $z_1$, the coordinate $z_{21} = z_2 - z_1$ lies in the limits $-z_1 \leq z_{21} \leq \ell - z_1$, see Figs. 7(b), (c). The Jacobian of the transformation $\{z_1, z_2\} \to \{z_1, z_{21}\}$ is of unity.

Step 5: Given an arbitrary function of the two variables $H(z_1, z_2)$, we can write out the identities for 2-fold integrals $(z_2 = z_1 + z_{21})$

$$I = \int_0^\ell dz_1 \int_0^\ell dz_2 H(z_1, z_2) = \int_0^\ell dz_1 \int_{-z_1}^{\ell-z_1} dz_{21} H(z_1, z_1 + z_{21}) =$$

$$\left( \int_{-\ell}^0 dz_{21} \int_{-z_{21}}^\ell dz_1 + \int_0^\ell dz_{21} \int_0^{\ell-z_{21}} dz_1 \right) H(z_1, z_1 + z_{21}), \tag{F1}$$

where, in the second line, we change the order of integration in accord with Fig. 7(c). It is easy to see that the two 2-fold integrals in Eq. (F1) are equal, i.e.

$$I = 2 \int_0^\ell dz_{21} \int_0^{\ell-z_{21}} dz_1 H(z_1, z_1 + z_{21}) \tag{F2}$$

if the function $H(z_1, z_2)$ depends on the arguments $\{z_1, z_2\}$ via the combinations $(z_1 + z_2)$ and $|z_{21}|$. These combinations are just the ones which enter into the Green function of Eq. (3.33c). To prove Eq. (F2) we replace the variable $z_{21}$ with $z_{21} = -\sigma_{21}$ in the first double integral of Eq. (F1), yielding

$$I_1 = \int_{-\ell}^0 dz_{21} \int_{-z_{21}}^\ell dz_1 H(z_1, z_1 + z_{21}) = \int_0^\ell d\sigma_{21} \int_{\sigma_{21}}^\ell dz_1 H(z_1, z_1 - \sigma_{21}),$$

and then replace $z_1$ with $z_1 = \sigma_1 + \sigma_{21}$, yielding finally

$$I_1 = \int_0^\ell d\sigma_{21} \int_0^{\ell-\sigma_{21}} d\sigma_1 H(\sigma_1 + \sigma_{21}, \sigma_1). \tag{F3}$$

In Eqs. (3.47a, c, d), the function $H(\sigma_1, \sigma_2)$ depends on the arguments $\{\sigma_1, \sigma_2\}$ via the combinations $(\sigma_1 + \sigma_2)$ and $|\sigma_2 - \sigma_1|$. Thus, the arguments of the function $H$ in Eq. (F3) are as follows $(\sigma_1 + \sigma_{21}) + \sigma_1 = 2\sigma_1 + \sigma_{21}$ and $|\sigma_1 - (\sigma_1 + \sigma_{21})| = |\sigma_{21}|$. The same combination of the through variables enters into the integrand $H(z_1, z_1 + z_{21})$ of the second 2-fold integral in Eq. (F1). Explicitly we have the following combination of the variables $z_1 + z_2 = 2z_1 + z_{21}$ and $|z_2 - z_1| = |z_{21}|$, hence, the second double integral of Eq. (F1) is equal to $I_1$ of Eq. (F3), Q.E.D.

Step 6: After these preliminaries, we calculate the integral in the expression $\overline{\upsilon}_{\Delta_1}$ of Eq. (3.47c)

$$\overline{\upsilon}_{\Delta_1} = \frac{1}{(A\ell)^2} \left( \frac{\ell}{A\xi_0^2 \overline{n}_t} \right) \int_0^{2\pi} d\varphi_1 \int_0^\ell dz_1 \int_{z_1}^{\sqrt{z_1^2 + (d/2)^2}} r_1 dr_1 \times$$

$$\int_{-\varphi_1}^{2\pi-\varphi_1} d\varphi_{21} \int_{-z_1}^{\ell-z_1} dz_{21} \int_{|z_{21}|}^{\sqrt{z_{21}^2+(d/2)^2}} r_{21}\, \upsilon(\vec{r}_1,\vec{r}_2) \bar{H}(z_1, z_1+z_{21}) dr_{21}, \tag{F4}$$

where the integrals over the variables $\{z_1, z_{21}\}$ are written out in the form similar to the first line of Eq. (F1), the function $\upsilon(\vec{r}_1,\vec{r}_2) = \gamma^2 \hbar^2 \tfrac{1}{2} r_{21}^{-3}\left(3(z_{21}/r_{21})^2 - 1\right)$ is due to Eq. (2.2), and the function $\bar{H}$ is Eq. (3.33c). The integrations over $\varphi_1$, $\varphi_{21}$, and $r_1$ give the factor $\tfrac{1}{2}(2\pi)^2 (d/2)^2$, the integral over $r_{21}$ of the function $r_{21}\upsilon(\vec{r}_1,\vec{r}_2)$ gives the factor $(d/2)^2 \left(z_{21}^2+(d/2)^2\right)^{-3/2}$. By bringing together all these factors, the integral $\bar{\upsilon}_{\Delta_1}$ in Eq. (F4) takes on the form of the 2-fold integral

$$\bar{\upsilon}_{\Delta_1} = \gamma^2\hbar^2 \frac{1}{\ell^2}\left(\frac{\ell}{A\xi_0^2 \bar{n}_t}\right)\int_0^\ell dz_1 \int_{-z_1}^{\ell-z_1} dz_{21}\left(z_{21}^2+(d/2)^2\right)^{-3/2} \bar{H}(z_1, z_1+z_{21}),$$

or by invoking Eq. (F2), we have

$$\bar{\upsilon}_{\Delta_1} = \gamma^2\hbar^2 \frac{2}{\ell^2}\left(\frac{\ell}{A\xi_0^2 \bar{n}_t}\right)\int_0^\ell dz_{21}\left(z_{21}^2+(d/2)^2\right)^{-3/2}\int_0^{\ell-z_{21}} dz_1 \bar{H}(z_1, z_1+z_{21}). \tag{F5}$$

If in Eq. (F5) we put $\bar{H}=1$, the integrand entering into the expression $\bar{\upsilon}_{\Delta_1}$ of Eq. (F5) is essentially the same as the integrand in the expressions $\bar{\upsilon}_0$ of Eq. (3.41a) and $\bar{\upsilon}_{\Delta_2}$ of Eq. (3.41d). From Eq. (F5) with $\bar{H}=1$, it follows

$$\bar{\upsilon}_0 = \gamma^2\hbar^2 \frac{2}{\ell^2}\int_0^\ell dz_{21}\left(z_{21}^2+(d/2)^2\right)^{-3/2}\int_0^{\ell-z_{21}} dz_1 = 2\pi\gamma^2\hbar^2/(A\ell), \tag{F6}$$

where $A = \pi d^2/4$ and the integral of Eq. (F6) is easily performed in the limit $\ell/d \to \infty$ by using the integral $\int_0^\infty (\cosh x)^{-2} dx = 1$ with $z_{21} = (d/2)\sinh x$. Given $\bar{\upsilon}_0$ of Eq. (F6) comparison $\bar{\upsilon}_0$ of Eq. (3.47a) with $\bar{\upsilon}_{\Delta_2}$ of Eq. (3.47d) gives us immediately

$$\bar{\upsilon}_{\Delta_2} = -2\pi\gamma^2\hbar^2/\left(A^2\xi_0^2 \bar{n}_t K^2\right). \tag{F7}$$

It remains to find the 2-fold integral $\bar{\upsilon}_{\Delta_1}$ of Eq. (F5) with the function $\bar{H}$ of Eq. (3.33c) which we rewrite as

$$\bar{H}(z_1, z_1+z_{21}) = (2K\sinh(K))^{-1}\left(\cosh\left(K\left\{1-\frac{2z_1+z_{21}}{\ell}\right\}\right) + \cosh\left(K\left\{1-\frac{|z_{21}|}{\ell}\right\}\right)\right). \tag{F8}$$

Relevant integral over $z_1$ in Eq. (F5) is performed invoking the integral $\int_0^h \cosh(\alpha+\beta z_1) dz_1 = \beta^{-1} \times (\sinh(\alpha+\beta h) - \sinh\alpha)$. The result reads

$$\bar{v}_{\Delta_1} = \gamma^2 \hbar^2 \frac{2}{\ell^2}\left(\frac{\ell}{A\xi_0^2 \bar{n}_t}\right)\int_0^\ell dz_{21}\left(z_{21}^2 + (d/2)^2\right)^{-3/2}\psi(z_{21}), \tag{F9}$$

where the function

$$\psi(z_{21}) = (2K\sinh(K))^{-1}\left((\ell/K)\sinh\left(K\left\{1-\frac{z_{21}}{\ell}\right\}\right) + (\ell - z_{21})\cosh\left(K\left\{1-\frac{|z_{21}|}{\ell}\right\}\right)\right). \tag{F10}$$

A decisive idea in handling the integral in Eqs. (F9), (F10) is making use the large parameter $\ell/d \to \infty$. In this a case, the integrand of Eqs. (F9), (F10) is the two-scale function. The function $\left(z_{21}^2 + (d/2)^2\right)^{-3/2}$ falls off to zero on the scale of fewer $d/2$ whereas the nonnegative monotone decreasing function $\psi(z_{21})$ falls off to zero on the interval $0 \leq z_{21} \leq \ell$ from $\psi(0) = \frac{1}{2}\ell\left(K^{-2} + K^{-1}\coth K\right)$, with $\psi(\ell) = 0$. It follows that asymptotically as $\ell/d \to \infty$, the function $\psi(z_{21})$ is essentially the constant $\psi(0)$ on the scale of fewer $d/2$ where the integrand of Eq. (F9) falls off to zero, yielding

$$\bar{v}_{\Delta_1} = \gamma^2 \hbar^2 \frac{2}{\ell^2}\left(\frac{\ell}{A\xi_0^2 \bar{n}_t}\right)\psi(0)\int_0^\ell dz_{21}\left(z_{21}^2 + (d/2)^2\right)^{-3/2} =$$

$$= \pi \frac{\gamma^2 \hbar^2}{(A\xi_0)^2 \bar{n}_t}\left(\frac{1}{K^2} + \frac{1}{K}\coth K\right). \tag{F11}$$

By gathering Eqs. (F11), (F7) we arrive at $\bar{v}_\Delta = \bar{v}_{\Delta_1} + \bar{v}_{\Delta_2}$ of Eq. (3.48b), in addition, $\bar{v}_0$ of Eq. (F6) is exactly Eq. (3.48a) of the main text.